\documentclass[aps,prb,twocolumn,preprintnumbers,amsmath,amssymb,floatfix]{revtex4}
\usepackage{graphicx, bm}
\usepackage{float}
\usepackage{latexsym}
\usepackage{amsmath}
\usepackage{graphics}
\usepackage{amssymb}
\usepackage{layout}
\usepackage{verbatim}
\usepackage{amsfonts,epsfig}
\usepackage{subfigure} 
\usepackage{dcolumn}
\usepackage{bm}
\usepackage{ulem}

\newcommand{\beq}{\begin{equation}}
\newcommand{\eeq}{\end{equation}}
\newcommand{\bk}{\bf k}
\begin{document}
\title{Temperature dependent compressibility in graphene and 2D systems}
\author{Qiuzi Li, E. H.\ Hwang, and S. Das Sarma}
\affiliation{Condensed Matter Theory Center, Department of Physics,
  University of Maryland, College Park, Maryland 20742}
\date{\today}

\begin{abstract}
We calculate the finite temperature compressibility for
two-dimensional semiconductor systems, monolayer graphene, and bilayer
graphene within the Hartree-Fock approximation.
We find that the calculated temperature dependent
compressibility including exchange energy is non-monotonic.
In 2D systems at low temperatures the  inverse compressibility decreases
first with increasing temperature, but after reaching a minimum it increases as
temperature is raised further. At high enough temperatures the negative
compressibility of low density systems induced by the exchange energy becomes
positive due to the dominance of the finite temperature kinetic
energy. The inverse compressibility in monolayer graphene is always positive
and its temperature dependence appears to be reverse of the 2D
semiconductor systems, i.e., it increases first with temperature and then decreases
at high temperatures. The  inverse compressibility of bilayer graphene
shows the same non-monotonic behavior as ordinary 2D systems, but at high
temperatures it approaches a constant which is smaller than the value
of the non-interacting bilayer graphene. We find the leading order temperature correction to the
compressibility within Hartree-Fock approximation to be $T^2 \ln T$ at
low temperatures for all three systems.
\end{abstract}

\maketitle

\section{Introduction}

In this article, we provide a detailed theory for the temperature-dependent electronic compressibility ($K$) of
an interacting two dimensional (2D) electron (or hole) system within
the Hartree-Fock theory. We consider three distinct 2D systems of
active current interest in condensed matter physics: 2D semiconductor
systems (e.g., quantum wells, heterostructures, inversion layers,
etc.); monolayer graphene; bilayer graphene. For the purpose of
comparison (and also for the sake of completeness), we also provide as
an appendix the corresponding finite temperature Hartree-Fock
compressibility for a standard three dimensional electron gas (3DEG)
since this result does not appear to be available in the theoretical
literature in spite of the sixty-year long history of studying
many-body interaction effects in 3DEG.

Ever since the pioneering measurement of the 2D compressibility by
Eisenstein {\it et al.}, \cite{eisenstein:1992,eisenstein:1994} it has
been extensively studied in 2D systems
because the compressibility provides fundamental insight into quantum ground state properties which are not readily obtained from transport measurements.
Some experimental studies \cite{dultz:2000,ilani:2000,DultzJiang_arXiv} of
the compressibility have shed light on
understanding the 2D metal-insulator transition
(MIT)\cite{dassarma:2005}.
The inverse compressibility
($K^{-1}$) is positive when the kinetic energy dominates over
interactions at high densities. As the 2D density is reduced  $K^{-1}$ changes
sign and becomes negative due to the increase of exchange energy
associated with electron-electron
interaction\cite{eisenstein:1992,eisenstein:1994}.
The negative $K^{-1}$ reaches a minimum value at a certain low
density $n$, and then
increases dramatically with further decreasing $n$ as the 2D MIT sets in.
The minimum point in $K^{-1}$  has occasionally been loosely identified as the critical
density for the 2D MIT and is closely related to the transition from
a homogeneous system to an inhomogeneous nonlinear screening
regime, \cite{ilani:2000,shi:2002} where the system is
dominated by the formation of electron (or hole) puddles. Whether this
observed low-density behavior of $K$ is a cause for or an effect of 2D
MIT is unclear. The low-density compressibility of 2D electronic (in this
paper we use the term ``electron" to imply either electron or hole as
should be obvious from the context) systems has been of interest for
almost 20 years now with the early experiments\cite{eisenstein:1992,eisenstein:1994,dultz:2000,ilani:2000,DultzJiang_arXiv} studying
semiconductor-based 2D systems extensively and very recent experiments
studying monolayer and bilayer graphene. The reason for the focus on
the density dependence of compressibility is that, in general, quantum
interaction effects increase monotonically with decreasing density in
a Coulomb system (with monolayer graphene being an odd exception where
the interaction parameter, the so-called graphene fine structure
constant, is independent of carrier density\cite{DasSarma_RMP11}), as discussed above,
leading eventually to the electronic compressibility becoming negative
at low, but easily accessible, densities due to exchange effects.

It is curious, however, that in spite of this intense interest in the strongly interacting low-density 2D compressibility, there has been little research on the theoretical functional dependence of 2D compressibility on temperature. This is strange because the scale for the temperature dependence of any electronic property is the Fermi temperature $T_F = E_F/k_B$, which invariably decreases with decreasing density. Thus, temperature effects on 2D compressibility become progressively more important {\it even at a fixed temperature} as the electron density goes down, since the important dimensionless temperature $t \equiv T/T_F$ increases with decreasing density even if $T$ is kept fixed. For example, 2D GaAs holes have $T_F < 1$ K for a hole density of $10^{10}$ cm$^{-2}$, which means that $T=1$ K is effectively a high-temperature regime for 2D hole densities $\lesssim 10^{10}$ cm$^{-2}$! Our work in this paper takes a first step in correcting this omission in the literature. Our results are important for interpreting low-density 2D compressibility (even at relatively low temperatures) existing in the literature.

The thermodynamic isothermal compressibility (or simply
compressibility) of a system is defined as
the change of pressure with volume, $K=-V^{-1}(\partial V/\partial P)_{N,T}$,
where $N$ is the particle number, $T$ is the temperature, $V$ is the
system volume,
$P=-(\partial E/\partial V)_{N,T}$ is the pressure, and $E$ is the
total energy of the system. Thus, $K^{-1} \equiv V (\partial^2 E/\partial V^2)$. In the non-degenerate or classical systems the velocity
of thermodynamic sound, $v_0$, often provides a convenient
experimental measure of compressibility,
$ v_0^2 = \frac{1}{m n}\frac{1}{K}$, where $m$ and $n=N/V$ are the mass
and the density of the system, respectively.
In the quantum limit the compressibility can be obtained theoretically by using the
theorem of Seitz \cite{pines:1966}, (i.e., by connecting $K$ with the change in chemical potential with
density, $\partial \mu /\partial n$, where
$\mu$ is the chemical potential and $n$ is the carrier density of the system).
Another method of evaluating $K$ is through the compressibility sum
rule. This is an exact relationship between the compressibility and
the long wavelength limit of the static dielectric function \cite{Mahan}.
In 2D quantum systems the compressibility is often measured from the
quantum capacitance \cite{eisenstein:1992,eisenstein:1994} which is
proportional to $dn/d\mu \propto K$.
A scanning single electron transistor has also been used to
directly measure the density dependent chemical potential \cite{ilani:2000},
 and hence $d \mu/dn$ or $dn/d \mu$. We will mostly discuss in this work the
 behavior of $K_0/K$, where $K_0$ is  the non-interacting compressibility, and $K_0/K \propto d \mu/dn$. We note that sometimes $dn/d \mu$ is referred to as the thermodynamic density of states since at $T=0$ and for noninteracting system $dn/d \mu \equiv D_0$, the noninteracting density of states.

In this paper, we refer to $K$ and $K^{-1}$ as compressibility and incompressibility (i.e. inverse compressibility) respectively, and also discuss results for  $d \mu/ dn$, which is sometimes directly measured experimentally, remembering that this quantity is directly proportional to the incompressibility. A part of our theoretical motivation for exploring the temperature dependence of
compressibility in 2D semiconductor systems arises from recent
experiments conducted on GaAs-based 2D systems \cite{gao:2011}. In addition,
there has been substantial experimental interest in the  temperature
dependence of graphene compressibility.\cite{Martin_NP08} We find that the calculated temperature dependent
compressibility including exchange energy is non-monotonic in temperature in both
graphene and 2D
semiconductor systems. However, their temperature behaviors appear to
be reversed, i.e.,
the inverse compressibility in graphene
$K^{-1} \propto d\mu /dn$ increases first with increasing temperature and then decreases
at high temperatures while the inverse compressibility in 2D semiconductor
systems decreases with temperature first and then increases at higher temperatures. We find that the
leading order temperature correction to 2D $K^{-1}$ within
Hartree-Fock approximation (HFA) is $t^2 \ln t$ at low temperatures
for both systems ($t \equiv T/T_F$). We also find that the BLG compressibility has
very weak temperature dependence because the non-interacting kinetic
energy is independent of temperature.


To calculate the compressibility we use Seitz's theorem which is
given by \cite{Mahan}
\begin{equation}
\frac{1}{K}=n^2 \partial\mu(T)/\partial n
\label{eq:def}
\end{equation}
where $\mu(T)$ is the finite temperature chemical potential and $n$ the
free carrier density.
In general, the direct measurement of $\partial \mu /\partial n$
provides information on
the thermodynamic many body
renormalization (for example, Fermi velocity)
arising from electron-electron interaction
effects\cite{Martin_NP08,Yafis_PRL07}.
Our goal here is to theoretically calculate the
renormalized $\partial \mu /\partial n$ in graphene and 2DEG including
exchange interaction effects, or equivalently in the HFA, which
should be an excellent quantitative approximation for compressibility in 2D systems.
Our calculated carrier density and temperature dependence of compressibility (or $\partial
\mu/\partial n$) can be
directly compared to experimental measurements in 2D systems including graphene.

The rest of the paper is organized as follows.
In Sec. \ref{sec:semi}, we introduce the formalism that will be used
in calculating compressibility in 2D semiconductor systems, which
includes the model describing the Coulomb system, the procedure of
calculation, and the numerical and analytical results of
compressibility in 2D semiconductor systems. In
Secs. \ref{sec:graphene} and \ref{sec:blg} we present the theoretical formalism and our
analytical and numerical results of compressibility in
MLG and BLG, respectively.
Sec. \ref{sec:discussion} contains discussions and
conclusions.

\section{Compressibility in 2D semiconductor systems}\label{sec:semi}

In this section, we present the compressibility of
2D semiconductor systems (2DS) such as Si
inversion layers in MOSFETs, 2D GaAs heterostructures  or quantum wells.
In some of our numerical calculations we use the parameters corresponding to
the n-GaAs or p-GaAs system, since these are the most-studied 2D system in the literature. When ever possible, we provide our results in dimensionless units for universal applicability.

In the absence of interaction, the non-interacting
finite temperature chemical potential, $\mu_0(T)$, is calculated through the
conservation of the total electron density, i.e.,
\begin{equation}
n(T)=n_0=\int d\epsilon D(\epsilon) n_F(\epsilon),
\label{eq:nt}
\end{equation}
where $n_0=n(T=0)$, $D(\epsilon)=gm/2\pi$ is the density of states
with $g$ being a degeneracy factor including both spin and possible valley degeneracy,
and $n_F(\epsilon) = [e^{\beta
    (\epsilon-\mu_0)}+1]^{-1}$ is the Fermi
distribution function. We take $\hbar =1$ throughout. By solving Eq.~(\ref{eq:nt}) self-consistently we have
the 2D chemical potential at finite temperatures
\begin{equation}
\mu_0(T)/E_F= t\ln(e^{1/t}-1)
\label{eq:noninter}
\end{equation}
where $E_F=k_F^2/2m = 2\pi n_0/gm$ is the Fermi energy at $T=0$ and $t
\equiv T/T_F$ with $T_F = E_F/k_B$.

To include interaction effects in the  chemical potential, we calculate the exchange self-energy contribution within HFA. We then have the interacting chemical potential as,
\begin{equation}
\mu = \mu_0 + \Sigma(k_F, E_F)
\label{eq:mu2d}
\end{equation}
where $\Sigma(k_F, E_F)$ is the HF exchange self-energy calculated at Fermi
momentum $k_F$. The HF exchange energy with the electron-electron
Coulomb interaction is given by\cite{Mahan}
\begin{equation}
\Sigma _{\text{ex}}(k)=-\int \frac{d^2k'}{(2\pi )^2}V(|k-k'|)n_F(\epsilon_{k'})
\label{eq:semiex}
\end{equation}
with $V(|k-k'|)=2\pi e^2/\kappa |k-k'|$ is the 2D bare Coulomb
interaction in the momentum space ($\kappa$ is the background lattice
dielectric constant), $\epsilon_{k}=k^2/2m$ is the single particle
energy ($m$ is the effective mass of the particle). In
the following, we calculate the compressibility of 2D
semiconductor systems for both zero temperature and finite
temperatures by finding the total chemical potential including exchange effects as a function of
density and temperature.

\subsection{Zero temperature compressibility}

At zero temperature ($T=0$),
the non-interacting part of the chemical potential is
just the Fermi energy $E_F$. Then $\partial\mu/\partial n$ becomes the inverse of the
noninteracting single-particle density
of states at the Fermi level: $(\partial n/\partial
\mu)=D_0(E_F)=gm/2\pi$, which is a density-independent constant for 2D.
From Eq.~(\ref{eq:def}) we have the zero temperature compressibility
of non-interacting 2D systems, $K_0$,
\begin{equation}
\frac{1}{K_0} = n^2 \frac{\partial \mu}{\partial n}=n^2/D_0.
\label{eq:K0}
\end{equation}
The noninteracting compressibility, $K_0$, is a positive quantity and
is inversely proportional to the square of particle density, $K_0 \propto n^{-2}$.

For interacting 2D systems the exchange self-energy can be calculated at $T = 0$ from
Eq. (\ref{eq:semiex}),
\begin{equation}
\Sigma _{\text{ex}}(k)=- \frac{2e^2}{\pi  \kappa }k_F E\Big(\frac{k}{k_F}\Big)
\label{eq:ex2D}
\end{equation}
where $E(x)$ is the the complete elliptic integral of the second kind.
The total chemical potential within HFA can be
calculated by putting $k=k_F$, i.e.,
\begin{equation}
\mu = \mu_0 + \Sigma(k_F) =  E_F (1 -\dfrac{2\sqrt{g}}{\pi}r_s ).
\label{eq:chemi}
\end{equation}
where  $g= g_v g_s$ ($g_s$, $g_v$ are the spin and valley degeneracy, respectively) and  the
dimensionless coupling constant $r_s=2 e^2 m/(\sqrt{g} \kappa k_F)$ is
the ratio of the average Coulomb potential energy $U$ to the average $T=0$
kinetic energy. The $r_s$ parameter as defined above is the dimensionless interaction parameter for a 2D electron liquid. We show most (except for Sec.~\ref{sec:semi}D) of our results in the dimensionless units of $t \equiv T/T_F$ and $r_s$, which would apply to any 2D semiconductor system. Differentiating
Eq.~(\ref{eq:chemi}) with respect to $n$ and making use of the relation
$k_F^2=4 \pi n/g$, we get the
zero temperature  inverse compressibility within HFA
\begin{equation}
\frac{K_0}{K} = 1-\frac{\sqrt{g}r_s}{\pi}
\label{eq:K0K}
\end{equation}
The interacting inverse compressibility $K_0/K$
monotonically decreases as $r_s$ increases (or density decreases) and
changes its sign from positive to negative at a
coupling strength $r_s = \pi/\sqrt{g}$.
This behavior has been
observed in experiments\cite{eisenstein:1992,eisenstein:1994} and
much discussed in the literature. Generally, negative compressibility leads to a thermodynamic instability of a system. However, the compressibility we have discussed in this paper applies only for the electronic part, i.e., the quantity
is not actually the compressibility of the whole system including the positive background charge
which is necessary for neutrality. The interaction between the electrons and its associated positive
neutralizing background has been ignored. In fact, a system with the negative electronic compressibility can be stabilized by the positive background, which gives rise to a positive compressibility for
the whole system. The negative electronic compressibility of the 2DEG has been directly measured
in experiments\cite{eisenstein:1992,eisenstein:1994,dultz:2000,ilani:2000,DultzJiang_arXiv} which probe only the electronic part of the compressibility, obviously the total
compressibility of such a system would still be positive in order to maintain thermodynamic stability
of the whole system.  Results given in
Eqs.~(\ref{eq:K0})--(\ref{eq:K0K}) are of course
well-known\cite{eisenstein:1994}, and are given here only for the sake
of completeness and for the sake of comparison with our finite-$T$
results discussed below.

\subsection{Finite temperature compressibility}
\label{results_A}

At finite temperatures
the chemical potential with exchange
self-energy can be calculated from Eq.~(\ref{eq:mu2d}) and
the normalized chemical potential, $\tilde{\mu}=\mu/E_F$ is expressed as
\begin{eqnarray}
\tilde{\mu}(T)=\tilde{\mu_0}(T) & - & \dfrac{2\sqrt{g}}{\pi}r_s \int _0^1\text{d}x
K_1(x)\Big[\dfrac{x}{e^{(x^2-\tilde{\mu _0})/t}+1} \nonumber \\
& + &\dfrac{1/x^2}{e^{(1/x^2-\tilde{\mu _0})/t}+1}\Big]
\label{eq:chemire}
\end{eqnarray}
where $\tilde{\mu_0}(T)=\mu_0/E_F$ is the non-interacting chemical
potential given in Eq.~(\ref{eq:noninter}), $t=T/T_F$, and
$K_1(x)$ is the complete elliptic integral of the first kind.

In the absence of exchange interaction, we obtain
the non-interacting
inverse compressibility at finite temperatures
\begin{equation}
\frac{1}{K_{ni}(T)}=n^2 \frac{\partial \mu_0}{\partial n}=\frac{h(t)}{K_0},
\label{eq:semiclass}
\end{equation}
where $h(t)=\frac{1}{1-e^{-1/t}}$ and $K_{ni}$ is the  finite
temperature non-interacting compressibility. For $T \ll T_F$, $h(t)
\sim 1 + \exp(-1/t)$. Thus we have $K_{ni}(T) \sim K_0(1-e^{-1/t})$. The leading
order correction to $K_{ni}(T)$ at low temperatures is exponentially suppressed.
For $T \gg T_F$, $h(t)\sim t$. Thus the noninteracting inverse
compressibility increases
linearly in the high temperature limit, $ {K_{ni}^{-1}(T\gg T_F) \sim t/K_0}$.

Including exchange effects, the finite temperature inverse compressibility
$K_0/K(T)$ can be obtained from Eq.~(\ref{eq:chemire})
\begin{widetext}
\begin{eqnarray}
\dfrac{K_0}{K(T)}=h(t) & - &
\frac{\sqrt{g}r_s}{\pi} \int _0^1\text{d}x
K_1(x)\Big[\dfrac{x}{e^{(x^2-\tilde{\mu _0})/t}+1}
  +\dfrac{1/x^2}{e^{(1/x^2-\tilde{\mu _0})/t}+1}\Big] \nonumber \\
& + & \frac{\sqrt{g}r_s}{2\pi t}\int _0^1\text{d}x K_1(x)\Big[\frac{x
    (x^2-h(t))}{\cosh ^2\frac{(x^2-\tilde{\mu}_0)}{2t}}
  +\frac{\frac{1}{x^2}(\frac{1}{x^2}-h(t))}{\cosh
    ^2\frac{(1/x^2-\tilde{\mu}_0)}{2t}}\Big].
\label{eq:comre}
\end{eqnarray}
\end{widetext}
The full numerical results of the compressibility are given in
Sec.~\ref{results_T}. We first consider the asymptotic behavior of the
temperature dependence both at high temperatures ($T\gg T_F$) and at low
temperatures ($T\ll T_F$).

In the low temperature limit ($t \ll 1$), the integrand in the
temperature-dependent term is only appreciable near the divergent
regime of the first elliptic function $K_1(x)$ ($x \approx 1 $). We
could get the asymptotic form by expanding the elliptic
integrals about that point, i.e.,
\begin{equation}
K_1(x \rightarrow 1) \approx
\ln\Big(\frac{4}{\sqrt{1-x^2}}\Big)+\frac{(1-x^2)}{4}
\Big[\ln\Big(\frac{4}{\sqrt{1-x^2}}\Big)-1\Big].
\label{eq:K1}
\end{equation}
Replacing $K_1(x)$ with Eq.~(\ref{eq:K1}) in the integrand
of Eq.~(\ref{eq:chemire}), the chemical potential in the low temperature
limit $T/T_F \ll 1$ is given by
\begin{equation}
\tilde{\mu}=\tilde{\mu_0}-\frac{2\sqrt{g}}{\pi}r_s [1+\frac{t^2}{8}(2
  A + B - A \ln\frac{16}{t})]
  \label{eq:2Dnormchem}
\end{equation}
where $A=\pi ^2/12$, $B=\frac{\pi
  ^2}{12}  [\ln (4\pi )+12\zeta '(-1)] \approx 0.449$ and $\zeta
'(-1)$ is the derivative of zeta function. Differentiating the
asymptotic formula of the many-body chemical potential, we obtain the
 inverse compressibility for $t\ll 1$,
\begin{eqnarray}
\begin{array}{l l l}
\dfrac{K_0}{K(T)}\Big|_{T \ll T_F} \simeq h(t)-\dfrac{\sqrt{g}}{\pi}r_s \Big[1-0.136 t^2
  -\dfrac{\pi^2}{32} t^2 \ln t\Big].
\end{array}
\label{eq:LHFSemi}
\end{eqnarray}
Since $h(t) \sim 1$ for $t \ll 1$, the leading order correction to the
inverse compressibility $K^{-1}(T)$ comes from the exchange energy
term, i.e., $t^2 \ln t$, which gives rise to the decrease of $K_0/K$ at low $T$ with
increasing temperature  as shown in
Fig.~\ref{fig:semi}(a). The asymptotic formula given in
Eq.~(\ref{eq:LHFSemi}) agrees well with our numerical
calculation at low temperatures as shown in Fig.~\ref{fig:semi}(a). The low-temperature behavior of $K^{-1}(T)$ is dominated by the $t^2 \ln t$ term which produces the shallow minimum in $K_0/K$ as a function of $t = T/T_F$ with the size of the minimum increasing with increasing (decreasing) $r_s$ (density). This interesting low-temperature non-monotonicity in the temperature-dependent 2D compressibility is entirely an exchange effect. We emphasize, however, that this exchange-induced minimum is very shallow in $K_0/K(T)$, and never exceeds $10\%$ of its $T=0$ value (often it is much less).

The asymptotic behavior of high temperature compressibility is
obtained by approximating the Fermi-Dirac distribution by the classical
Boltzmann distribution. For $t \rightarrow \infty$, the normalized
chemical potential with exchange energy is given by
\begin{equation}
\tilde{\mu}(T)=\tilde{\mu_0}-\frac{\sqrt{g \pi}}{2 \sqrt{t}}r_s.
\end{equation}
The corresponding asymptotic formula for the high temperature inverse compressibility is given by
\begin{equation}
\dfrac{K_0}{K(T)} \Big|_{T \gg T_F} \approx h(t)-\sqrt{\dfrac{\pi}{2t}}r_s.
\label{eq:semiht}
\end{equation}
For $t \gg 1 $, the exchange energy contribution to the  inverse compressibility decreases as $t^{-1/2}$ while the kinetic energy
contribution increases lineally since $h(t)\sim t$. Thus, the kinetic
term dominates in the high temperature limit and the combined  compressibility approaches the non-interacting result as it should at very high temperatures. At high
temperatures the role of the exchange-correlation effects is
diminished due to the increase of the thermal kinetic energy. As a consequence,
the negative compressibility at low densities becomes positive at
a high enough temperature, $T=T_F(\pi r_s^2/2)^{1/3}$. This
 behavior (reversing sign due to increasing temperature) was observed in high mobility p-GaAs
systems\cite{shapira:1996}. We have explicitly verified that our numerical HFA results agree precisely with the asymptotic high-temperature result of Eq. \ref{eq:semiht} for $t \gg 1$. We note that the exchange correction to $K_0/K$ falls off very slowly only as $r_s/\sqrt{t}$ for $t \gg 1$, and as such quantum effects are quite large at large $r_s$ even for $T \gg T_F$. In fact, the classical regime in compressibility is approached only for $T \gg r_s^2 T_F$.

\subsection{Numerical results of compressibility in 2D semiconductors}
\label{results_T}

\begin{figure}
\begin{center}
\includegraphics[width=0.7\linewidth]{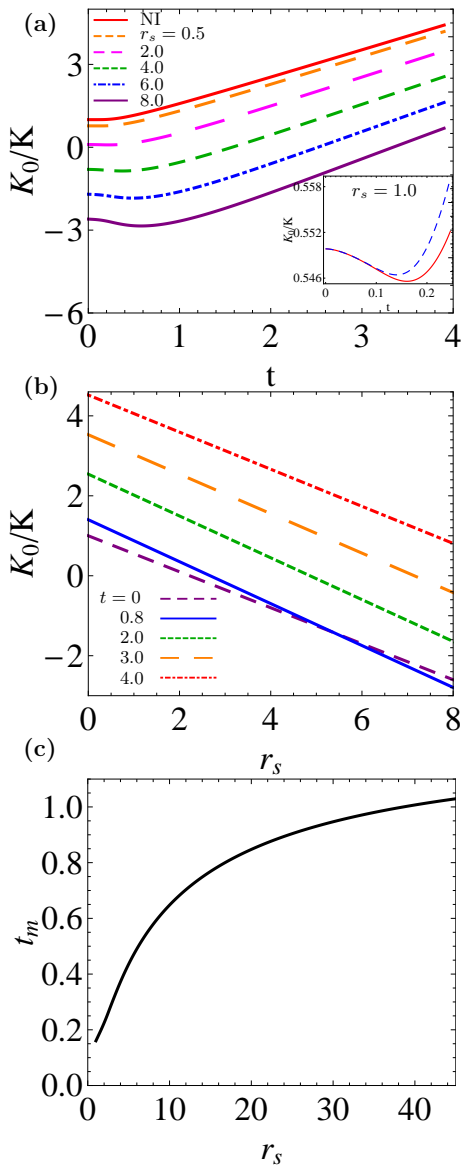}
\caption{(color online) (a) Calculated $K_0/K$ in 2DEG as a function
  of rescaled temperature $t$ for different values of  $r_s$. The inset presents $K_0/K$ in the low temperature regime. The dashed and solid lines are corresponding to the analytical results (given in Eq.~\ref{eq:LHFSemi}) and the numerical results, respectively. (b)
  The  inverse compressibility $K_0/K$ in 2DEG as a function of $r_s$  for
  different values of rescaled temperature $t$. (c) Calculated $t_m$,  where the minimum of $K_0/K$ occurs, as a function of $r_s$. }
  \label{fig:semi}
\end{center}
\end{figure}

In Fig.~\ref{fig:semi}(a), we show the calculated inverse
compressibility $K_0/K$  as a function
of rescaled temperature $t$ for six different values of $r_s$.
The red solid line corresponds
to the non-interacting case ($K_{ni}$) without the exchange
interaction. As shown in Sec.~\ref{results_A} $K_{ni}^{-1}(T) \sim h(t) \sim t$ for $t \rightarrow \infty$ increases
linearly in the high temperature classical limit.
For finite $r_s$ the zero temperature inverse compressibility
decreases with $r_s$ and eventually becomes negative (if
$r_s > \pi/\sqrt{2}$) due to the exchange energy.
As shown in Fig.~\ref{fig:semi}(a)  $K_0/K$ is not
a monotonic function of rescaled temperature $t$.
As $t$ increases from zero $K^{-1}(t)$ initially decreases since the kinetic
energy term is exponentially suppressed due to the Fermi surface
restriction, but the exchange term keeps decreasing as $t^2 \ln t$.
After reaching the minimum value at an $r_s$-dependent characteristic
temperature $t_m$, $K^{-1}$ increases linearly at high temperatures.
This non-monotonic behavior becomes stronger at higher $r_s$. More
interestingly the negative value of $K^{-1}$ at low temperature and high $r_s$
reverses its sign as the temperature increases. At high enough
temperatures, where the kinetic energy dominates over the interaction
energy, the system always has a positive compressibility, but typically strongly suppressed in magnitude from $K_{ni}^{-1}$ except for $t \rightarrow \infty$. Thus, many-body effects manifest strongly in the compressibility even for $T > T_F$ in most situations.

\begin{figure}
\begin{center}
\includegraphics[width=0.7\linewidth]{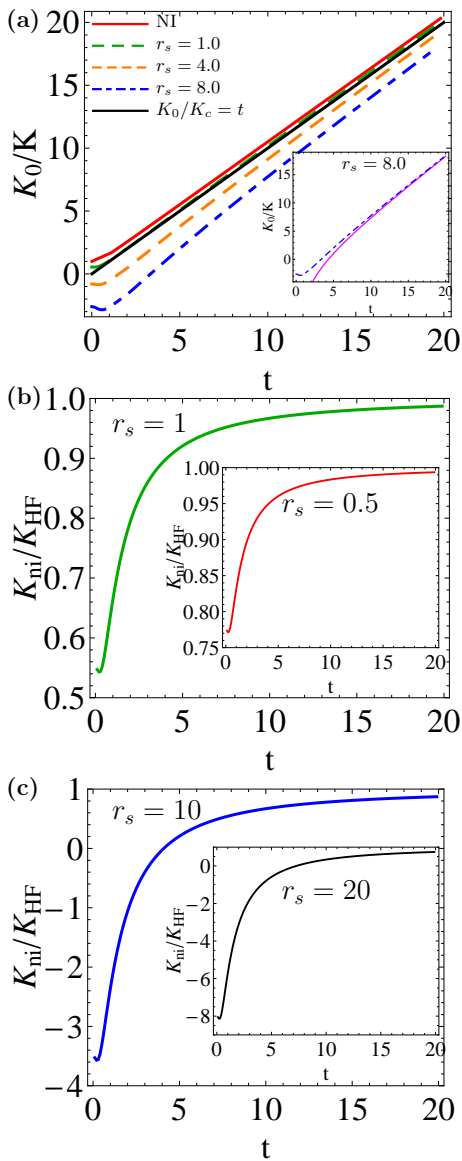}
\caption{(color online) (a) Calculated $K_0/K$ as a function of $t$
  for $r_s = 1, 4, 8$ comparing the HF result with the non-interacting
  result $K_0/K_{ni}=h(t)$ as in Eq.~\ref{eq:semiclass} and the pure
  classical result $K_0/K_c=t$. Inset shows the HF result and the asymptotic form for the high
  temperature $K_0/K$ as in Eq. \ref{eq:semiht} with $r_s = 8$. (b)
  and (c) show $K_{ni}/K_{HF}$ for $r_s = 0.5$, $1.0$, $10$, $20$
  for $t=0\sim 20$. }
  \label{fig:semi2}
\end{center}
\end{figure}

In Fig.~\ref{fig:semi}(b), we show our numerically calculated  inverse compressibility $K_0/K$ as a
function of $r_s$ for different values of the rescaled temperature
$t$. It is clear that the interacting $K_0/K$ manifests
monotonically stronger many-body effects with increasing $r_s$. The
curves for $t=0$ and $t=0.8$ cross each other at finite value of $r_s$,
which corresponds to the non-monotonic temperature dependence of
inverse compressibility $K_0/K$ in the low temperature regime. As an inset in Fig.~\ref{fig:semi}(a) we explicitly show a quantitative comparison between our derived low-$T$ asymptotic formula (Eq.~\ref{eq:LHFSemi}) and the exact numerical HFA results verifying the presence of the $t^2 \ln t$ term, which leads to the low-$t$ minimum, in the HF $K_0/K$. In Fig.~\ref{fig:semi}(c), we show as a function of $r_s$ our numerically calculated value of $t(\equiv T/T_F)=t_m$ where the low-$t$ minimum of $K_0/K$ occurs. We note that although $t_m$ increases monotonically with increasing (decreasing) $r_s$ (density), this increase is sublinear implying that the actual temperature (in Kelvin) $T_m$, where the minimum occurs, decreases (increases) with decreasing (increasing) carrier density $n$ since $T_F \propto n$ and $r_s \propto n^{-1/2}$. This means that the experimental observation of this non-monotonicity of $K_0/K$ as a function of temperature may be extremely difficult, if not impossible, with the experimental $K_0/K$ manifesting only a monotonic increase with increasing temperature at all densities. The experimental observation is further hampered by two additional complications: (1) the actual decrease in the magnitude of $K_0/K$ associated with the shallow minimum is rather small ($< 10\%$); (2) the low-density, large-$r_s$ regime of the 2D system often develops strong disorder-driven density inhomogeneity.

In Fig.~\ref{fig:semi2} we compare our numerical results with our high-temperature analytic theory as given in Eq.~\ref{eq:semiht}. The purely classical compressibility $K_c(T)$ is given by $K_0/K_c =t$, which is the $t(=T/T_F) \rightarrow \infty$ limit of $K_0/K_{ni}\equiv h(t)= (1-e^{-1/t})^{-1}$, as given in Eq.~\ref{eq:semiclass}. We note that the exchange correction to the non-interacting result, as given in Eq.~\ref{eq:semiht} falls off very slowly as $r_s t^{-1/2} \sim 1/\sqrt{T}$. In Fig.~\ref{fig:semi2}(a) we show calculated $K_0/K$ in the $t=0-20$ range for $r_s = 1, 4, 8$ comparing the HF result with the non-interacting result $K_0/K_{ni}=h(t)$ as in Eq.~\ref{eq:semiclass}, and the pure classical result $K_0/K_c=t$. For the sake of comparison, we also show (for $r_s=8$) the corresponding high-temperature HFA result as in Eq.~\ref{eq:semiht}. The interesting point to note here is that, as can be seen from Eq.~\ref{eq:semiht}, the quantum exchange correction is quantitatively substantial even at a temperature $T \gg T_F$ for large $r_s$ with the HFA results being quantitatively well below (by $\sim 20 \%$) the classical result even for $T = 20 T_F$. Although this appears somewhat counter-intuitive, the importance of quantum interaction persisting to high temperatures ($T \gg T_F$) for large $r_s$ can be understood by considering the relative magnitudes of the three dimensionless energy parameters $r_s \equiv U/E_F$, $t \equiv T/T_F$, and $\Gamma \equiv U/k_B T$ that control the physics of an interacting quantum system -- we note that $r_s$, $t$, $\Gamma$ are not independent parameters since $t \equiv r_s/\Gamma$. The classical noninteracting limit requires both $t$, $\Gamma^{-1} \gg 1$ which necessitates $t \gg r_s$ as well. Thus one condition for the classical limit is $t \gg r_s$ when $r_s >1$ (and $t\gg 1$ for $r_s <1$), which is much stronger than $t \gg 1$ when $r_s$ is large! In Fig.~\ref{fig:semi2}(b) and (c) we show $K_{ni}/K_{HF}$ for $r_s = 0.5$, $1.0$, $10$, $20$ for $t=0\sim 20$ to emphasize that the exchange correction to the  inverse compressibility is substantial in magnitude even for $T \gg T_F$.

\subsection{Results for 2D GaAs systems}
Since there has been considerable experimental activity\cite{eisenstein:1992,eisenstein:1994,Millard_PRB97,shapira:1996} in measuring the 2D compressibility of both electrons\cite{eisenstein:1992,eisenstein:1994,Millard_PRB97} and holes\cite{shapira:1996} in GaAs-based two-dimensional semiconductor systems, we provide in this section a set of numerical results for the HF compressibility of 2D GaAs systems (electrons and holes) as a function of carrier density ($n$ in the unit of cm$^{-2}$) and temperature ($T$ in K). The parameters used in these numerical calculations are for electrons (holes): $m=0.07 (0.4) m_e$; $g=2$ (i.e., $g_s =2$, $g_v=1$); $\kappa=13$, where $m_e$ is the free electron mass in vacuum and $\kappa$ is the background lattice dielectric constant of GaAs-AlGaAs heterostructure. There is no valley degeneracy ($g_v = 1$) in GaAs, and we consider the spin-degenerate ($g_s =2$) zero-magnetic field situation.

\begin{figure}
\begin{center}
\includegraphics[width=0.99\linewidth]{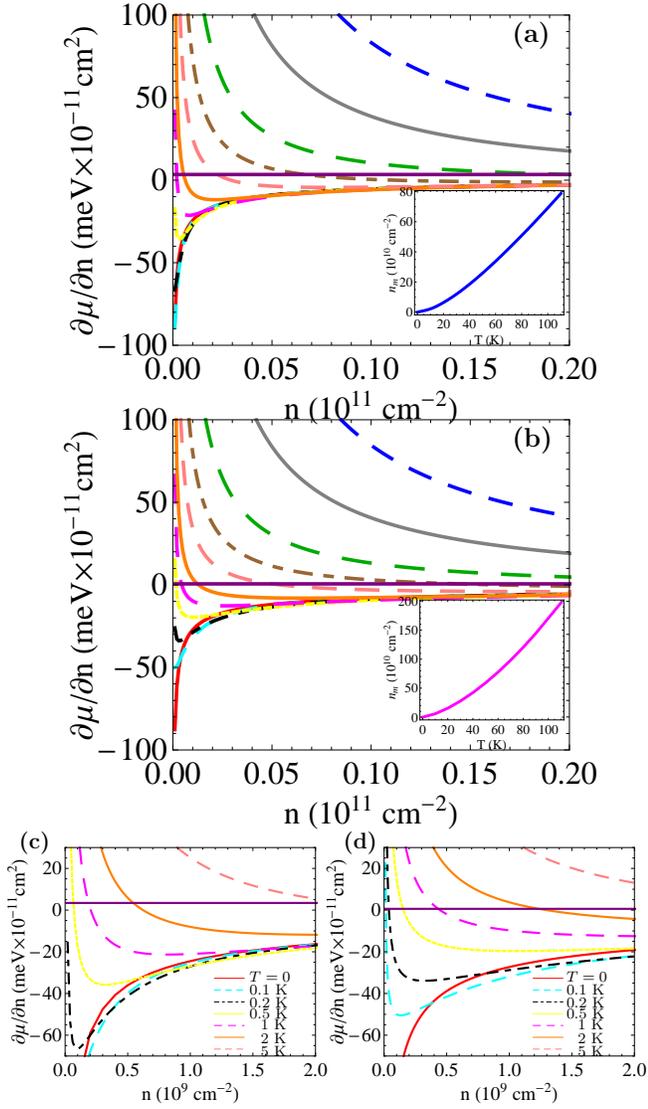}
\caption{(color online) (a) and (b) Calculated $d\mu/dn \equiv (K
  n^2)^{-1}$ in the HFA for 2D GaAs electrons ($m=0.07 m_e$) and holes
  ($m=0.4 m_e$) respectively for $T=0, 0.1, 0.2, 0.5, 1, 2, 5, 10, 20,
  50, 100$ K (from bottom to top). The solid horizontal line corresponds to the
  non-interacting $T=0$ result, $d \mu/dn \equiv [D_0(E_F)]^{-1}$. Insets of
  (a) and (b): Calculated $n_m$,  where the minimum of $d \mu/dn$
  occurs, as a function of $T$ for electrons and holes respectively.
  (c) and (d) Zoom-in figures for the low-density parts of $d\mu/d n$
  versus $n$ for 2D GaAs electrons and holes respectively. }
  \label{fig:semi3}
\end{center}
\end{figure}

In Fig.~\ref{fig:semi3}(a) and (b) we show our calculated $d\mu/dn
\equiv (K n^2)^{-1}$ in the HFA for 2D GaAs electrons and holes
respectively for $T=0, 0.1, 0.2, 0.5, 1, 2, 5, 10, 20, 50, 100$ K. The
corresponding non-interacting $T=0$ result, $d \mu/dn \equiv
[D_0(E_F)]^{-1}=0.24(m_e/m)10^{-11}$ meVcm$^2$, is also shown as a
constant horizontal line in each
figure. The calculated temperature dependence is stronger for the
holes than the electrons in Fig.~\ref{fig:semi3} since the scale for
the $T$-dependence, $T_F$, is much smaller (almost by a factor of six for the
same density) for the holes compared with the electrons because of the
large difference in the two effective masses ($0.07$ versus
$0.4$). For $n=2 \times 10^{10}$ cm$^{-2}$: $T^e_F=8.3$ K, $T^h_F=1.5$ K. Of course the qualitative behavior of $K(n,T)$ is the same for both electrons and holes, it is only that the temperature scale for the holes is lower.

The qualitative behavior of $d\mu/d n$ as a function density ($n$) and temperature ($T$) as shown in Fig. \ref{fig:semi3} is in very good agreement with experimental results\cite{shapira:1996}, showing that the temperature dependence of compressibility can indeed be very important in samples with low densities (and consequently with low $T_F$), particularly for holes because of their large effective mass. We note that $T_F = 41.5$ K (electrons) and $7.3$ K (holes) for a density of $n=10^{11}$ cm$^{-2}$ (and $T_F \propto n$).

A particular qualitative feature of Fig.~\ref{fig:semi3} deserves
special attention: $d\mu/d n(\propto K^{-1})$ as a
function of $n$ for a fixed temperature shows a
very sharp upward turn with well-defined minimum at low densities,
particularly for the low-$T$ results. This striking low-$n$ (and
low-$T$) non-monotonicity arises from a quantum-classical crossover
effect which turns on when $T_F(n)<T$ from being $T_F >T$ at some density
and can therefore only be seen for low-$T$ and low-$n$ results. The
minimum occurs at a density $n_m$ which we plot as a function of
temperature $T$ in the inset showing that $n_m$ increases with
increasing $T$.
To further emphasize this interesting behavior, we
show in Fig.~\ref{fig:semi3}(c) and (d), the low-density parts of
$d\mu/d n$ versus $n$ for a few low $T$ values where the minimum is
clearly visible. This sharp increase of $K^{-1}$ or $d\mu/d n$ for low
density (and low temperature) has been experimentally observed in 2D
GaAs systems, and has often been associated with the 2D
metal-insulator-transition (MIT) driven by disorder. By contrast, our
theory does not include disorder effects, we only include
finite-temperature effects which are, of course, very strong at low
densities where $T_F(n)<T$. The intuition based on the $T=0$ theory of
compressibility clearly must fail at some low density (i.e., for
$T_F(n)\lesssim T$) since the $T=0$ limit of $K^{-1}$ becomes large
(in magnitude) and negative at very low $n$ whereas the finite-$T$
theory ({\it for any finite $T$}) predicts that at the lowest
densities, where $T_F(n)<T$, $K^{-1}$ must become large and positive
(i.e., the classical behavior for $T \gg T_F$), in fact eventually
diverging as $T/T_F$ for $T \gg T_F$! We believe, based on the results
presented in Fig.~\ref{fig:semi3}, that experimentalists should
re-investigate the older data for $K^{-1}$ at low densities where
finite temperature effects may be playing a significant role. There
are some recent experimental results Ref.[\onlinecite{gao:2011}]
supporting our finding, but more experimental data are necessary in the interacting low density and low temperature regime to settle this question definitively.

\begin{figure}
\begin{center}
\includegraphics[width=0.99\linewidth]{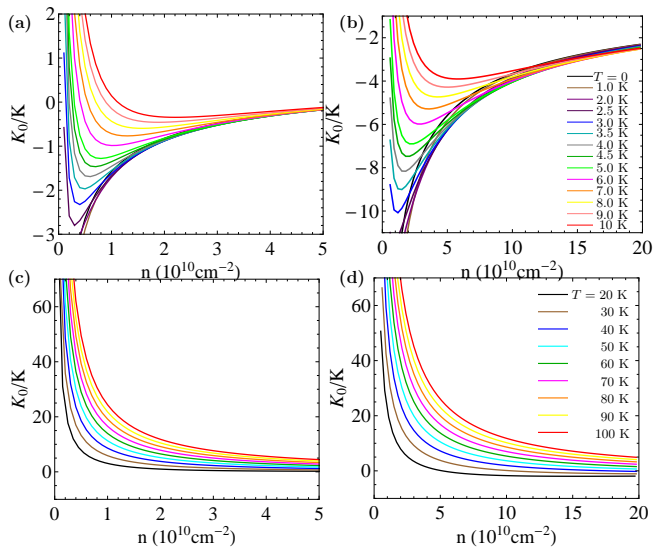}
\caption{(color online) (a) and (b) Calculated $K_0/K$ as a function
  of carrier density in the HFA for
  electrons ($m=0.07 m_e$) and holes ($m=0.4 m_e$), respectively, over a
  wide range of temperature $T=0-10$ K. (c) and
  (d) show the high-$T$
  behavior ($T=20-100$ K) of $K_0/K$ for electrons and holes
  respectively. The temperature increases from bottom to top in each
  figure.}
  \label{fig:semi4}
\end{center}
\end{figure}

Given the possible qualitative importance of finite temperature corrections to the 2D compressibility as discussed above (and shown in Fig.~\ref{fig:semi3}), we provide in Fig.~\ref{fig:semi4} the calculated $K_0/K$ in the HFA for both electrons (Fig.~\ref{fig:semi4}(a)) and holes (Fig.~\ref{fig:semi4}(b)) over a wide range of density and temperature. The non-monotonicity apparent for higher-$T$ results in Figs.~\ref{fig:semi4}(a) and (b) particularly for the hole data, is actually present in all the curves except that the non-monotonicity manifests itself at much lower density than the range covered in Fig.~\ref{fig:semi4} for the curves that simply look like that $K_0/K$ keeps on decreasing with density monotonically. In Figs.~\ref{fig:semi4}(c) and (d) we show the high-$T$ behavior ($T=20-100$ K) of $K_0/K$.

\begin{figure}
\begin{center}
\includegraphics[width=0.99\linewidth]{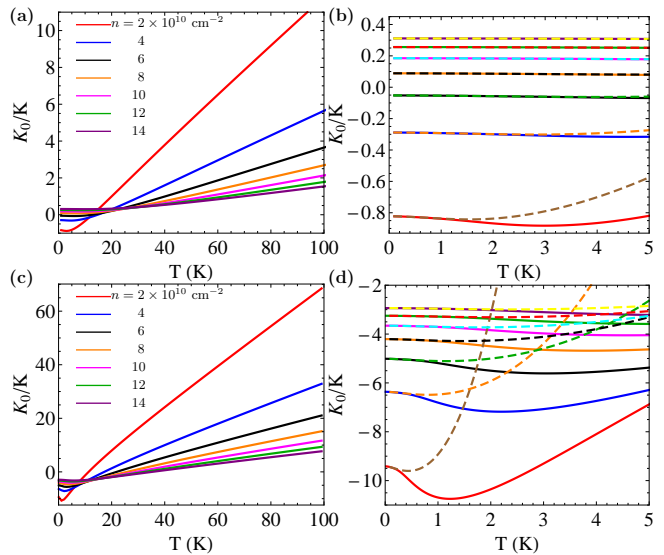}
\caption{(color online) Calculated $K_0/K$ in the HFA as a function of temperature for various densities. Figs.~(a) and (b) are for electrons($m=0.07 m_e$), (c) and (d) are for holes($m=0.4 m_e$). (a) and (c) show $K_0/K$ in the HFA for
  a wide range of temperature ($T= 0-100$ K) for various
  densities $n=2, \ 4, \ 6,\ 8, \ 10, \ 12, \ 14 \times 10^{10}$
  cm$^{-2}$ (from top to  bottom). (b) and (d) show $K_0/K$ at low temperatures ($T=0-5$ K) for various
  densities $n=2, \ 4, \ 6,\ 8, \ 10, \ 12, \ 14 \times 10^{10}$
  cm$^{-2}$ (from bottom to top) and compare the
  low-T analytical results (dashed lines) and numerical results (solid lines).
}
  \label{fig:semi5}
\end{center}
\end{figure}

In Fig.~\ref{fig:semi5}, we show our calculated compressibility for 2D
GaAs electrons and holes as a function of temperature for a few
densities. Figs.~\ref{fig:semi5}(a) and (b) correspond to electrons
showing $K_0/K$ as a function of temperature respectively over a wide
$T=0-100$ K range [\ref{fig:semi5}(a)] and a narrow low temperature
range $T=0-5$ K [\ref{fig:semi5}(b)] whereas
Figs.~\ref{fig:semi5}(c) and (d) show the same for 2D holes in GaAs. In
Figs.~\ref{fig:semi5}(b) and (d), we show a comparison between our
low-$T$ analytical and numerical results also. The important point to
note in Fig. \ref{fig:semi5} is that the low-temperature minimum
should be observable in 2D semiconductor systems in careful
measurements of $K_0/K$ as a function of temperature provided disorder
effects are unimportant, i.e., highest-mobility samples are used for
the study. Lower density samples would typically manifest deeper
minima as can be seen in Fig.~\ref{fig:semi5}.

\section{Compressibility in graphene}\label{sec:graphene}

In this section, we theoretically calculate the finite temperature
compressibility of monolayer graphene including exchange
interaction effects. It has been shown that the HFA is an
excellent quantitative approximation due to the small contribution of
the correlation energy in graphene\cite{Euyheon_PRL07}. We focus on extrinsic graphene, i.e., gated or doped graphene with a tunable
2D free carrier density $n$ of electrons (holes) in the conduction
(valence) band, i.e., the chemical potential $\mu$ being positive
(negative). For undoped (intrinsic) graphene the chemical potential is zero even at finite temperatures, which gives rise to a logarithmically divergent
compressibility if disorder effects are neglected \cite{Martin_NP08,DavidHwang_PRB11}. The basic feature of graphene compressibility and its functional dependence on density have been well studied in the literature at $T=0$\cite{Euyheon_PRL07,DavidHwang_PRB11}.

We use the same definition for the compressibility
given in Sec. \ref{sec:semi}.
However, there are two main differences between the usual 2D
semiconductor and graphene. The
first one is the difference between their energy dispersion relation: 2D semiconductor has parabolic dispersion relation while
graphene has linear dispersion. The kinetic energy in
graphene is given by $\epsilon _{k,s}=s v_F|k|$, where
$|k|$ is the wave vector with respect to the Dirac point, $v_F$ is the Fermi velocity with the value of $v_F \sim 10^8 $ cm/s and $s=\pm 1$
for the conduction and valence band respectively.  The second
difference is that we need to consider the contribution from valence
band electrons in graphene because graphene is a gapless semiconductor so that the valence band and the conduction band
touch  each other at the Dirac point.
In graphene the total degeneracy $g=4$ (from spin degeneracy $g_s=2$
and valley degeneracy $g_v=2$).

Since the non-interacting chemical
potential of graphene is $\mu_0=E_F= \hbar v_F k_F$ at zero temperature the
corresponding non-interacting  inverse  compressibility is given by
\begin{equation}
K_0^{-1}=n^2 \pi \hbar v_F /(2 k_F),
\end{equation}
where we use $g=4$ and the electron density is related to the Fermi
wave vector as $k_F^2= \pi n$. Unlike the regular 2D systems,
noninteracting $d\mu/ dn$ in graphene
is density dependent (i.e. $d \mu/dn \propto
n^{-1/2}$) due to the linear energy dispersion of graphene.

The chemical potential within HFA is the sum of the
non-interacting kinetic energy part and the exchange self-energy part. The exchange
self-energy for graphene is given as\cite{Euyheon_PRL07}:
\begin{equation}
\Sigma _{x,s}(k)=-\sum _{s'} \int \frac{d^2k'}{(2\pi )^2}V\left(\left|k-k'\right|\right)n_F\left(\xi _{\bm k',s'}\right)F_{s,s'}\left(\bm k,\bm k'\right)
\label{eq:gex}
\end{equation}
where $s,s'=\pm 1$ indicates the band indices, $V\left(q\right) = 2 \pi e^2
/ (\kappa q)$ is the bare Coulomb interaction ($\kappa$ is the
background dielectric constant), and $F_{ss'}(\bm
k,\bm k')=(1 + ss'\cos\theta_{\bm k\bm k'})/2$ arises from the
wavefunction overlap factor where $\theta_{\bm k\bm k'}$ is the angle
between $\bm k$ and $\bm k'$. In regular (i.e., non-chiral) 2D systems, $F\equiv 1$, but in graphene, which is a chiral material, the chiral factor $F$ is important due to the underlying pseudospin dynamics. With Eq.~(\ref{eq:gex}) we calculate the
interacting  compressibility of graphene at finite temperatures. Since
the zero temperature exchange self energy can be calculated
analytically\cite{Euyheon_PRL07}
we first consider the zero temperature compressibility as the starting point of our discussion.

\subsection{Zero temperature compressibility}

At $T=0$, the Fermi distribution function in Eq. \ref{eq:gex} becomes $n_F
(\xi) = \theta(E_F-\xi)$.
We divide the exchange self-energy into two
parts\cite{Euyheon_PRL07}: one from the intraband transition
$\Sigma^{\rm int}_{x}$, and the other from the interband transition
$\Sigma^{\rm ext}_{x}$. That is,
$\Sigma_{{\rm x},s}(k) = \Sigma_{{\rm x},s}^{\rm int}(k) +
\Sigma_{{\rm x},s}^{\rm ext}(k)$, where
\begin{equation}
\begin{array}{l l l }
\Sigma_{{\rm x},s}^{\rm int}(k) =  -\int \dfrac{d \bm q}{(2\pi)^2}
V_c(\bm q)\, F_{s,-}(\bm k, \bm k-\bm q);
\\
\\
\Sigma_{{\rm x},s}^{\rm ext}(k) = -\sum_{s'}\int \dfrac{d \bm q}{(2\pi)^2}
\delta n_F(\xi_{\bm k - \bm q,s'})
V_c(\bm q)\,F_{ss'}(\bm k, \bm
k-\bm q)
\end{array}
\end{equation}
where $\delta n_F(\xi_{\bm k - \bm q,s'}) = n_F(\xi_{\bm k- \bm q,s'})
- \frac{1}{2}(1 - s')$ is the difference in the electron occupation
from the intrinsic $T=0$ case.

After some algebra (the detail of the
derivation is given in Ref.~[\onlinecite{Euyheon_PRL07}]), we find
the density dependent exchange contribution to the inverse compressibility
$K_0/K$ at zero temperature
\begin{equation}
\frac{K_0}{K}=1+r_s\Big[\dfrac{1}{4}   \ln
  \Big(\dfrac{4}{y}\Big)-\dfrac{3}{8}-\frac{1}{2 \pi} -\frac{C}{ \pi} \Big]
\end{equation}
where $y=k_F/k_c$ with a momentum cut-off $k_c \sim a^{-1}$ where $a$
is a lattice constant of graphene,
$C \approx 0.916$ is
Catalan's constant, and $r_s=\dfrac{e^2}{\hbar  v_F  \kappa }$ is
the graphene coupling constant (or fine structure constant). We note
that the $\ln y \equiv \ln (k_F/k_c)$ term takes care of the divergent
compressibility of intrinsic graphene which does not enter our
discussion in any significant manner. Unlike in 2D systems, where the
interaction parameter $r_s \sim n^{-1/2}$, for graphene $r_s$ is a constant in
density due to its linear energy dispersion.
However, by adjusting the background dielectric constant ($\kappa$) we can vary
the $r_s$ value, from $r_s = 2.2$ (for $\kappa =1$, graphene suspended in vacuum) to very small $r_s$ by making $\kappa$ very large. For graphene on SiO$_2$, $r_s = 0.8$.  Note that unlike ordinary 2D systems the calculated graphene
compressibility with the exchange correction is always positive, which
has recently been measured by several different techniques recently \cite{eisenstein:1992,eisenstein:1994}.

\subsection{Finite temperature compressibility}

In this subsection, we present the theoretical formalism of
the finite temperature $K_0/K$  and its
asymptotic analytical formula  at  low temperatures ($t
\ll 1$). We consider extrinsic graphene with $E_F >0$ and concentrate on the situation with the chemical potential lying in the conduction band with no loss of generality.

The finite temperature chemical potential (without exchange energy) $\mu
_0$  must be calculated by the conservation of the total electron
density, i.e.  $n(T) = n(T=0) + p(T)$, where $n(T)$ and $p(T)$ are the
electron and hole  density at $T$, respectively. They are given by
\begin{equation}
n(T) = g\int \frac{d^2p}{(2\pi)^2} \frac{1}{\exp[\beta(\varepsilon_p -\mu)] +
1}
\end{equation}
and
\begin{equation}
p(T) = g\int \frac{d^2p}{(2\pi)^2} \frac{1}{\exp[-\beta(\varepsilon_p -\mu)] +
1},
\end{equation}
where $g$ is the total degeneracy and $\beta = 1/k_BT$.
Thus, we have the self consistent equation for $\mu$ as
\begin{equation}
\frac{1}{2} \left (\frac{T_F}{T} \right )^2 = F_1\left (\beta{\mu} \right ) -
F_1 \left (-\beta{\mu} \right ),
\end{equation}
where $F_n(x)$ is given by
\begin{equation}
F_n(x) = \int_0^{\infty} \frac{t^n dt}{1+\exp(t-x)}.
\end{equation}
Then we obtain the non-interacting chemical potential for both low and
high temperature
limits for graphene as
\begin{eqnarray}
\mu_0(T) & \approx & E_F\left [ 1-\frac{\pi^2}{6} \left( \frac{T}{T_F} \right
  )^2 \right
  ] \;\; {\rm for } \; T/T_F \ll 1 \label{eq:chel} \\
\mu_0(T) & \approx & \frac{E_F}{4 \ln 2} \frac{T_F}{T}
\hspace{2.2cm}  {\rm for}
\; T/T_F \gg 1.
\label{eq:chet}
\end{eqnarray}
With Eqs.~(\ref{eq:chel}) and (\ref{eq:chet}) we can find
the noninteracting graphene compressibility at low temperatures and at high
temperatures, respectively. Let
$\mu_0/E_F \equiv \tilde{\mu_0}$ and $t=T/T_F$, then the
non-interacting part of the inverse compressibility
can be written as
\begin{eqnarray}
\frac{K_0}{K_{ni}(T)} &= & 1+ \frac{\pi^2 t^2}{6} \hspace{1.cm} {\rm for } \;
t \ll 1,  \label{eq:graNI1}\\
\frac{K_0}{K_{ni}(T)} & = & \frac{1}{2 \ln 2} \frac{1}{t}
\;\hspace{1.cm} {\rm for} \; t \gg 1.
\label{eq:graNI2}
\end{eqnarray}
Thus the leading order correction
to the inverse compressibility $K_{ni}(T)^{-1}$ increases quadratically at
low temperatures and then it decreases inverse linearly at high
temperatures. This fact indicates that the graphene compressibility
shows the opposite behavior to the ordinary 2D systems as shown in
Sec.~\ref{sec:semi}, where the inverse compressibility of 2D systems
decreases first and then increases as the temperature increases. We emphasize that this qualitative difference in the noninteracting finite-$T$ compressibility between 2D systems and monolayer graphene arises both from the linear, chiral energy dispersion of graphene and from its gaplessness allowing for thermal excitations of conduction valence band electron-hole pairs at finite $T$.

Now we calculate the full compressibility at low temperatures,
including the exchange energy.
As done for the zero temperature case, the exchange self energy
of graphene can be separated into contributions from the intrinsic
part $\Sigma^{\rm int}_{x}$ and the extrinsic part
$\Sigma^{\rm ext}_{x}$\cite{Euyheon_PRB09}.  We can see that the first
term in Eq.~(\ref{eq:E}) is $\Sigma^{\rm int}_{x}$ while the second term
is $\Sigma^{\rm ext}_{x}$. For $\mu>0$  the Hartree-Fock self
energy in graphene at finite temperature becomes
\begin{eqnarray}
\begin{array}{l c c c c c c}
\Sigma^{\rm int}_{x}(k,T) =-\dfrac{e^2}{2\pi  \kappa }\int\int
\frac{q dq d\theta}{|k-q|}\frac{1}{\exp \left(\beta \left(-\hbar  v_F q-\mu
  \right)\right)+1}\frac{1- \text{cos$\theta $}}{2}
\\
\\
\Sigma^{\rm ext}_{x}(k,T)= -\dfrac{e^2}{2\pi  \kappa }\int \int
\frac{qdq d\theta}{|k-q|}\frac{1}{\exp \left(\beta \left(\hbar  v_F q-\mu_0
  \right)\right)+1}\frac{1+ \text{cos$\theta $}}{2}
\label{eq:E}
\end{array}
\end{eqnarray}
The intrinsic exchange self-energy
$\Sigma^{\rm int}_{x}$ in Eq.~(\ref{eq:E}) can be derived by using the
fact that the exponential term in
$\Sigma^{\rm int}_{x}$ is exponentially suppressed and almost equal to
$1$ in the low temperature limit. Then, we have \cite{Euyheon_PRL07}
\begin{equation}
\begin{array}{l l l }
\Sigma^{\rm int}_{x}(T)  = \dfrac{e^2 k_F}{\pi  \kappa } \left [ -
  \dfrac{E(y)}{y}+
  \dfrac{\pi }{4}\ln (\dfrac{4}{y})-\dfrac{\pi }{8} \right .  \\
\ \ \ \ \ \ \ \ \ \ \ - \left . \int _0^y\dfrac{dx}{x^3} ( K_1(x)-E(x)-\dfrac{\pi
  }{4}x^2) \right ] + O(e^{-\beta}),
\end{array}
\end{equation}
where $y=k_F/k_c \ll 1$, $K_1(x)$ and $E(x)$ are the complete
elliptic integral of the first and second kinds, respectively.
Note that the leading order temperature correction of the intrinsic exchange
self-energy at low temperatures is exponentially suppressed. On the
other hand, the extrinsic part of the exchange self-energy becomes
\begin{equation}
\Sigma^{\rm ext}_{x}(T) = -\dfrac{e^2 k_F}{\pi  \kappa }
\left [ C+\dfrac{1}{2}-\dfrac{B_g}{12} \pi ^2 t^2
+\dfrac{1}{12} \pi ^2 t^2 \ln t \right ]
\label{eq:2}
\end{equation}
where $C \approx 0.916$ is Catalan's constant, $B_g=-1+2 \gamma -12 \ln (G)+\ln \left(\frac{128}{\pi
}\right)\approx 0.877$ where $G\simeq 1.282$  is Glaisher's constant and
$\gamma\simeq 0.577$ is Euler's constant. Differentiating the total chemical
potential with respect to the
density, we have the asymptotic form of  inverse
compressibility in the low temperature limit
\begin{eqnarray}
\dfrac{K_0}{K(T)}=f(t) & + & r_s\Big[\dfrac{1}{4} \ln
  \Big(\dfrac{4}{y}\Big)-\dfrac{3 }{8}-\dfrac{C}{\pi}-\dfrac{1}{2 \pi} \nonumber \\
& - & \dfrac{\pi t^2}{12}  (B_g-1)+\frac{\pi}{12}  t^2
\ln  t\Big],
\label{eq:comreg}
\end{eqnarray}
where $f(t) \equiv 1+ \frac{\pi^2 t^2}{6}$ is the non-interacting part
of the inverse compressibility given in Eq.~(\ref{eq:graNI1}).
In the low temperature limit the leading order temperature correction
to the total  inverse  compressibility is the same as that in the 2D
semiconductor system due to the exchange energy, i.e., $t^2 \ln
t$. However, the logarithmic correction can be detected only at very low temperatures ($t<0.01$). In general, the $t^2$ correction from the non-interacting part dominates at low temperatures ($t>0.01$) and the
positive coefficient of $t^2$
term gives rise to increasing behavior of $K^{-1}$ with temperature.
Our asymptotic results at
low temperatures agree very well with the full numerical results shown in Fig.~\ref{fig:gra}(a). The results provided in this section generalize the existing graphene literature on compressibility\cite{Euyheon_PRL07} to finite temperatures.

\subsection{Numerical results of graphene compressibility}

\begin{figure}
\begin{center}
\includegraphics[width=1.0\linewidth]{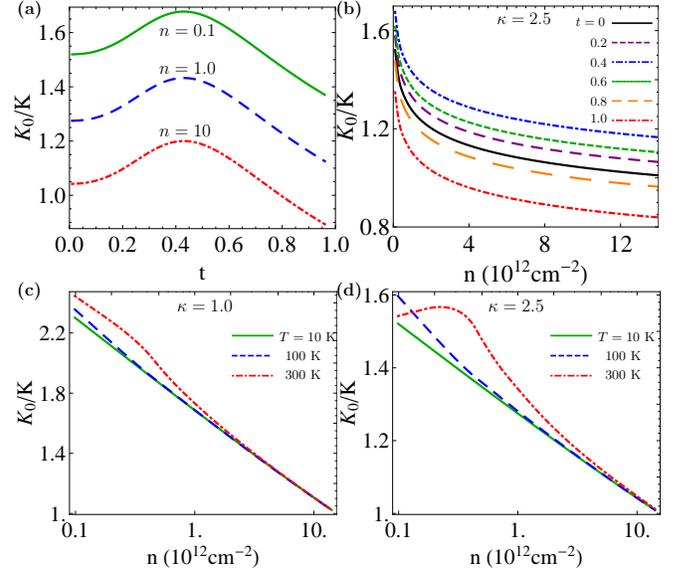}
\caption{(color online). (a) Calculated $K_0/K$ in graphene  as a function of temperature
  $t$ for three different values of density in the units of
  $10^{12}$cm$^{-2}$. (b) Calculated $K_0/K$ in graphene  as a function of carrier
  density $n$ for six different values of rescaled temperature
  $t$. In this calculation the momentum cut-off $k_c=1/a$ ($a=2.46$
  {\AA}) and  $\kappa=2.5$ are used. (c) The semi-log plot of $K_0/K$
  versus carrier density with different temperatures $T$ for
  $\kappa=1$ and (d) for $\kappa=2.5$. }
  \label{fig:gra}
  \end{center}
\end{figure}

In this subsection, we present our numerical results of finite
temperature compressibility in graphene.
Throughout this section, we use $v_F = 10^8 $ cm/s and the wave
vector cut-off $k_c =1/a$ ($a=2.46$ {\AA}). We show results for
suspended graphene ($\kappa \sim 1.0$) and for graphene on SiO$_2$
($\kappa \sim 2.5$). In Fig.~\ref{fig:gra}(a) we
show the temperature dependence of the total  inverse compressibility $K_0/K$
for three different values of carrier density $n=0.1\times10^{12}
cm^{-2}$, $n=1\times10^{12}$ cm$^{-2}$, $n=10.0\times10^{12}$
cm$^{-2}$ with dielectric constant
$\kappa=2.5$ (corresponding to graphene on the Si/SiO$_2$ substrate). We can clearly see that the temperature dependence is
non-monotonic and this non-monotonic behavior mostly comes from the
non-interacting compressibility, which is different from the 2D semiconductor
case. Unlike the 2D
parabolic-band case, the compressibility in graphene does not change
sign in the range of experimentally relevant parameters and typically exchange corrections are always quantitatively small ($\lesssim 20 \%$). We show the
carrier density dependence of $K_0/K$ in Fig.~\ref{fig:gra}(b), where
the $K_0/K$ is a monotonically decreasing function of carrier density
for small values of $t$ ($T\simeq 400K$ for $t=1$ and
$n=0.1\times10^{12} cm^{-2}$). In Fig.~\ref{fig:gra}(c) and (d), we present the calculated $K_0/K$ as
a function of carrier density for different temperatures. For fixed
temperatures, $K_0/K$ is a decreasing function of $n$ at higher
carrier density, which corresponds to the low temperature limit manifesting $n^{-1}$ density dependence. On the other hand, at lower carrier
density, $K_0/K$ has $n^{1/2}$ density dependence, which corresponds
to the high temperature limit and the main contribution to $K_0/K$ comes from the non-interacting part (see Eq.~\ref{eq:graNI2}).

\begin{figure}
\begin{center}
\includegraphics[width=0.99\linewidth]{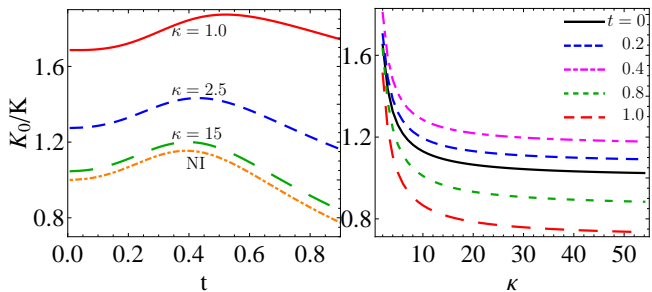}
\caption{(color online). (a) Calculated $K_0/K$ in monolayer graphene as a function of temperature $t$
  for three different values of background dielectric constant within
  HFA for carrier density $n=10^{12}$cm$^{-2}$. The dot-dashed (orange) line presents the temperature dependence of the non-interacting $K_0/K$. (b) Calculated $K_0/K$ in monolayer graphene as a function of the background dielectric constant $\kappa$ for a few fixed values of $t$. $k_c=1/a$ ($a=2.46${\AA}). }
  \label{fig:gracom}
  \end{center}
\end{figure}

In Fig.~\ref{fig:gracom}, we compare the temperature dependence for
different values of background dielectric constant $\kappa=1$
(i.e. graphene in vacuum), $\kappa=2.5$ (graphene on Si/SiO$_2$) and
$\kappa=15$ (graphene on hafnium oxide HfO$_2$), respectively. The
value of $r_s$ represents the fine structure constant of graphene which depends only on $\kappa$ and
does not depend on the carrier density because of the linear energy dispersion. Larger dielectric constant
corresponds to smaller values of $r_s$, which indicates a weak-coupling
system in terms of electron-electron
interaction\cite{Euyheon_PRB09}. Note that the trends of  $K_0/K$
among different values of dielectric constant are similar. The larger
the value of dielectric constant (the smaller the value of $r_s$), the
less strong is the density dependence of $K_0/K$. We also show in this
figure our calculated $K_0/K$ in HFA
compared with the noninteracting $K_0/K$
as a function of the background dielectric constant $\kappa$ for a few fixed values of $t$. We could see that the inverse
compressibility $K_0/K$ first increases as the temperature $t$
increases, reaches a maximum, and then decreases as the temperature
$t$ further goes
up. This non-monotonic behavior of $K_0/K$ dependence on the
temperature is mainly due to the non-interacting part of the chemical
potential, which is consistent with the calculation of the non-interacting
part of the chemical potential (Eq.\ref{eq:chet}). From Eq~(\ref{eq:chet}) we know that the temperature dependence of the chemical potential is $t^2$
(with a positive slope) at low temperatures while the temperature
dependence of the chemical potential approximates as $1/t$ (with a negative slope) at higher temperatures. There must be an intermediate point at which the chemical potential changes its sign of slope and so does the compressibility $K$. Since $K^{-1}$ is the derivative of chemical
potential with respect to $n$, it has similar temperature
dependence as the chemical potential. Therefore, the non-interacting
part of $K_0/K$ behaves as $t^2$ at low temperatures and then
changes to $1/t$ behavior. The domination of graphene compressibility by noninteracting effects is just a direct manifestation of the weakness of electron-electron interaction in graphene compared with the 2D semiconductor case.


\section{compressibility of BLG}
\label{sec:blg}

In this section, we calculate the BLG compressibility within the
HF approximation.
There is some related theoretical work in the
literature dealing with the problem of compressibility in BLG,
\cite{kusminskiy:2008,borghi:2010,DavidHwang_PRB11,DMHD_arXiv11}
but the specific and detailed temperature dependent compressibility
presented in the current work is not available in the literature.
To calculate BLG compressibility at finite temperatures  we use the two-band
approximation, which is valid for the low density limit. At high density, BLG dispersion approaches the linear
energy dispersion of monolayer graphene as the Fermi energy becomes large\cite{DasSarma_RMP11}, and therefore the high-density BLG compressibility behaves similar to the MLG compressibility studied in the last section.
It has been shown that the two band model presents the
most important qualitative signatures of the BLG compressibility.
\cite{kusminskiy:2008,borghi:2010,Sensarma_PRB11,DavidHwang_PRB11}
For low density bilayer graphene within two-band model, the
quadratic approximation is commonly used, i.e., $\xi_k = \hbar^2 v_F^2
k^2/\gamma_1 = \hbar^2 k^2/(2m)$, where $\gamma_1$
is the interlayer hopping parameter from the tight-binding approximation and
$m = \gamma_1/(2v_F^2)$ is the effective mass of the electrons. This dispersion
comes from a low-energy effective theory of bilayer graphene\cite{DasSarma_RMP11}, which
essentially discards the two split bands and confines electrons to
those lattice sites not involved in the interlayer coupling.
While the quadratic
dispersion is relevant at low energies, the actual dispersion of the
bands is hyperbolic. At large wave vectors, relevant at large
densities, the dispersion is effectively linear and the system should
behave like single layer graphene. Thus our results obtained in the
previous section should be applicable to the high density BLG.

The exchange self energy of BLG is given by
\begin{equation}
\Sigma_{x,s}(k) = - \sum_{s'}\sum_{{\bf k}'}V_{{\bf
    k-k}'}n_F(\xi_{{\bf k}',s'})F_{s,s'}({\bk},{\bk}'),
\label{eq:BLG}
\end{equation}
where $V_{\bk-\bk'} = 2\pi e^2 /\kappa |\bk-\bk'|$ is the Coulomb
potential, $\xi_{\bk}=k^2/2m$, $n_F(x)$ is the Fermi distribution
function, and
$F_{s,s'}({\bk},{\bk'})=(1+ss'\cos(2\theta_{\bk\bk'}))/2$ is the wave
function overlap factor and $\theta_{kk'}$ is the angle between $\bk$
and $\bk'$. To get the explicit temperature dependence we rewrite
Eq.~(\ref{eq:BLG}) as
\begin{eqnarray}
\Sigma_{x}(k) & =& \Sigma_{2D}(k) + \Sigma_x^{int}(k) \nonumber \\
& + &
\frac{1}{2}\sum_{\bk'}V_{\bk-\bk'} [
  f(\xi_{\bk'})+ g(\xi_{\bk'})] [1-\cos(2\theta)],
\label{eq:BLG2}
\end{eqnarray}
where $f(x) = 1/(e^{\beta(x-\mu)} + 1)$ and $g(x)=1/(e^{\beta(x+\mu)}+1)$,
$\Sigma_{2D}(k)$ is the ordinary 2D self energy given in Eq.~(5), and
$\Sigma_x^{int}$ is given by
\begin{eqnarray}
\Sigma_x^{int}(k) &=& -\int \frac{d^2 k'}{(2\pi)^2}V_{k-k'} \frac{1-\cos(2\theta)}{2} \\
&=&\frac{2 e^2 k}{3\pi  \kappa } \Big[\frac{\pi }{2}-p(z)
+q(z)\Big] \nonumber
\end{eqnarray}
where
\begin{align}
p(z) =\begin{cases}
\dfrac{(1+7 z^2)E(z)}{3 z^3} & z \leq 1\\
\\
\dfrac{(1+7 z^2)E(1/z)}{3 z^2} & z > 1 \end{cases}
\end{align}
 and
 \begin{align}
q(z) =\begin{cases}
\dfrac{(1+2 z^2 -3 z^4)K_1(z)}{3 z^3} & z \leq 1\\
\\
\dfrac{4(z^2-1)K_1(1/z)}{3 z^2} & z > 1 \end{cases}
\end{align}
Here $z= k/k_c$, $k_c \sim 1/a$ is the momentum cut-off (similar to the MLG situation) with a
 lattice constant $a$. $K_1(x)$ and $E(x)$ are the complete elliptic
 integral of the first and the second kind, respectively.
Since we have studied $\Sigma_{2D}$ in Sec.~II and $\Sigma_x^{int}$ is
independent of temperature,  the new feature of temperature dependent
compressibility of BLG arises from
the third term in Eq.~(\ref{eq:BLG2}), i.e.,
\begin{equation}
\Sigma_x^{(3)}(k)  =  \frac{1}{2}\sum_{k'}V_{k-k'} [
  f(\xi_{k'})+ g(\xi_{k'})] [1-\cos(2\theta)].
\label{eq:BLG3term}
\end{equation}
Then the total temperature dependent HFA chemical potential of BLG can be
calculated to be
\begin{equation}
\mu(T) = E_F + \Sigma(k_F,T).
\end{equation}

\subsection{Zero temperature BLG compressibility}

At $T=0$, using $f(x) = \theta(1-x)$ and $g(x) =0$ in
Eq.~(\ref{eq:BLG3term}), the exchange
self-energy can be calculated as:
\begin{eqnarray}
\Sigma_{x}(k) &=& - \frac{2e^2 k_F}{\pi  \kappa }
E\left(\frac{k}{k_F}\right) + \frac{2 e^2 k}{3\pi  \kappa }
\Big[p\left(\frac{k}{k_F}\right)-q\left(\frac{k}{k_F}\right)
\nonumber  \\
&-&p\left(\frac{k}{k_c}\right)+q\left(\frac{k}{k_c}\right)\Big].
\end{eqnarray}
Then, the exchange self-energy at $k=k_F$ is given by
\begin{equation}
\Sigma_{x}(k_F) = - \frac{2e^2 k_F}{\pi  \kappa }+\frac{2 e^2
  k_F}{3\pi  \kappa } \Big[\frac{8}{3}
-p(y)+q(y)\Big].
\end{equation}
where $y=k_F/k_c$. The total chemical potential for BLG within HFA is then written as
\begin{equation}
\mu = E_F \left[1-\frac{r_s}{\pi}\left(\frac{2}{9}+\frac{2}{3}[p(y)-q(y)]\right)\right]
\label{eq:chblg}
\end{equation}
where $r_s = 2 e^2 m / (\kappa k_F)$ is the dimensionless BLG coupling. In the following numerical calculation, we use the the value of
dielectric constant $\kappa= 2.5$ and effective mass $m=0.03
m_e$\cite{DasSarma_RMP11}. Differentiating Eq.~\ref{eq:chblg} with respect to $n$ and
making use of the relation $k_F^2 = \pi n$, we get the zero
temperature BLG inverse compressibility within HFA
\begin{equation}
\frac{K_0}{K} = 1-\frac{r_s}{\pi}\left(\frac{1}{9}+\frac{1}{3}[p(y)-q(y)]\right)+\frac{r_s}{\pi} w(y)
\label{eq:blgcomzero}
\end{equation}
where $K_0$ is the non-interacting compressibility of BLG, given by
\begin{equation}
\frac{1}{K_0}=\frac{\pi n^2}{2m}.
\label{eq:BLGzero}
\end{equation}
and
 \begin{align}
w(y) =\begin{cases}
\dfrac{1 + y^2}{3y^3} E(y) +\dfrac{ y^2-1}{3y^3}  K_1(y) & y \leq 1\\
\\
\dfrac{1 + y^2}{3y^2} E(1/y) +\dfrac{1- y^2}{3y^2}  K_1(1/y) & y > 1 \end{cases}.
\end{align}
At low enough densities the zero temperature compressibility within HF
approximation becomes negative similar to the corresponding 2D
semiconductor system. The corresponding density is $n_c \sim 10^{10}$ $cm^{-2}$ or $r_s \sim 27$, for which the negative compressibility
manifests itself. This is a very small (large) value of density
($r_s$) compared with the
ordinary 2D case where the negative compressibility shows up for $r_s
> 2.2$. However the negative
compressibility obtained in HFA has not been experimentally observed in bilayer graphene
systems\cite{henriksen:2010,young:2010,martin:2010}.
The non-negative compressibility may be attributed to correlation effects.
It is shown that the contributions
from exchange and correlation almost cancel each other out in BLG at low
density, leaving a
small positive compressibility at all densities in contrast to the corresponding 2DEG systems \cite{eisenstein:1992,eisenstein:1994,dultz:2000,ilani:2000,DultzJiang_arXiv}
where the electronic compressibility can become negative at low densities. \cite{borghi:2010,Sensarma_PRB11}
The non-quadratic band structure also gives rise to
significant affects on the compressibility as has recently been discussed in the literature. \cite{DavidHwang_PRB11}

\begin{figure}
\begin{center}
\includegraphics[width=0.7\linewidth]{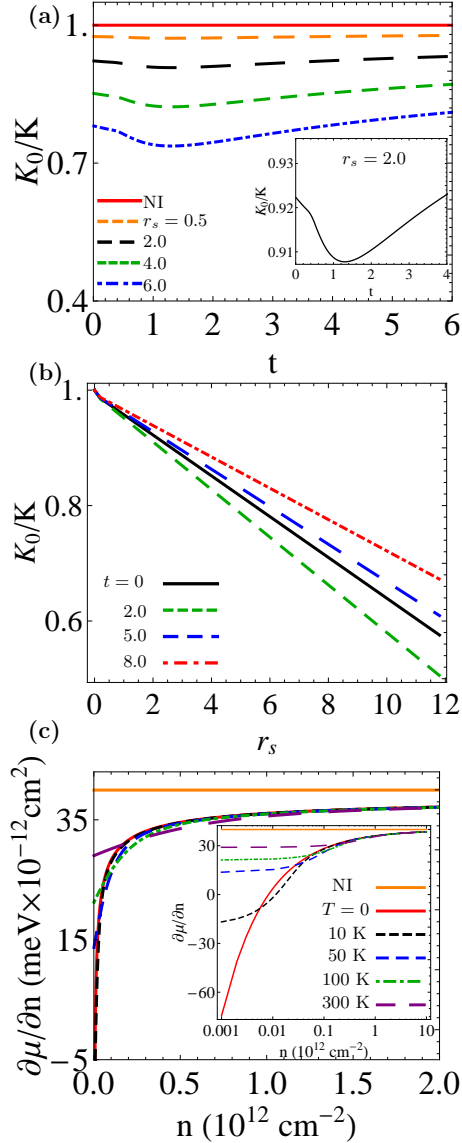}
\caption{(color online) (a) Calculated $K_0/K$ in BLG as a function
  of temperature $t=T/T_F$ for different values of  $r_s$. The
  inset presents $K_0/K$ for $r_s = 2.0$ showing the non-monotonic
  behavior. $k_c=1/a$ ($a=2.46$ {\AA}) and $\kappa=2.5$
   are used. (b) The  inverse
  compressibility $K_0/K$ in BLG as a function of $r_s$  for
  different values of temperature $t$. (c) $\partial \mu
  /\partial n$ in BLG versus carrier density $n$ for
   various temperatures. Inset shows the details in the low density limit
   in the semi-log scale.}
  \label{fig:blg}
\end{center}
\end{figure}

\subsection{Finite temperature BLG compressibility}

At finite temperatures the non-interacting chemical potential is
temperature independent in BLG, i.e.,
\begin{equation}
\mu_0(T) = E_F.
\end{equation}
Then the non-interacting compressibility of BLG is also temperature
independent and is the same as the zero temperature compressibility
given in Eq.~(\ref{eq:BLGzero})
\begin{equation}
{K_{ni}(T)}=K_0.
\end{equation}
At finite temperatures, the temperature dependence of BLG
compressibility  comes from the exchange contribution. This is very
intriguing because the non-interacting kinetic energy dominates exchange
self energy at high temperatures for both ordinary 2D systems
and MLG. But in BLG the entire temperature dependence arises from the
exchange energy! Thus, any experimentally observed temperature dependence in the BLG compressibility must entirely be a many-body effect, at least within the 2-band approximation (i.e., at not-too-high carrier densities).

Let us consider the exchange self energy,
Eq.~(\ref{eq:BLG2}), at finite temperatures.
The first term in Eq.~(\ref{eq:BLG2})
has been calculated in Sec.~\ref{sec:semi} [see Eq.~(\ref{eq:2Dnormchem})]
and the second term is temperature independent.
The third term [Eq.~(\ref{eq:BLG3term})] becomes in the low
temperature limit $t=T/T_F \ll 1$:
\begin{equation}
\Sigma_x^{(3)}(k_F) \simeq \frac{2 e^2 k_F}{3\pi  \kappa }
(\frac{8}{3}-\frac{\pi }{2}-\frac{\pi^2 }{24} t^2) +O(t^4).
\end{equation}
Thus, we have the total HFA chemical potential in the low temperature limit
\begin{eqnarray}
\mu(T)&=&E_F \left[1-\frac{r_s}{\pi}\left(\frac{2}{9}+\frac{2}{3}[p(y)-q(y)]\right)\right] \nonumber \\
&-&E_F\frac{r_s}{\pi}\left[\frac{B t^2}{4}+\frac{5 \pi ^2
    t^2}{72}-\frac{1}{48} \pi ^2 t^2 \text{ln}\frac{16}{t}\right].
\label{eq:BLGnormchem}
\end{eqnarray}
where $B=\frac{\pi
  ^2}{12}  [\ln (4\pi )+12\zeta '(-1)] \approx 0.449$ and $\zeta
'(-1)$ is the derivative of the zeta function. Differentiating the
asymptotic formula of chemical potential with respect to the density,
we have the BLG $K_0/K$ for $t\ll 1$
\begin{eqnarray}
\begin{array}{l l l}
\dfrac{K_0}{K(T)} \simeq \dfrac{K_0}{K(T=0)}+\dfrac{r_s}{\pi} \Big[0.55 t^2+\dfrac{\pi^2}{32}  t^2 \text{ln}t\Big].
\end{array}
\label{eq:LHFBLG}
\end{eqnarray}
where $K_0/K(T=0)$ has been derived in Eq.~\ref{eq:blgcomzero}.
We have the same leading order term $t^2\ln t$ arising from the exchange self-energy as the ordinary 2D
system. Thus, at low temperatures the calculated inverse compressibility decreases
as the temperature increases.

In the high temperature limit $T\gg T_F$, the chemical potential can be calculated to be:
\begin{equation}
\mu (T)= E_F \left[1-\frac{r_s}{\pi}\left(\frac{B_1}{\sqrt{t}}+\frac{2}{3}[p(y)-q(y)]\right)\right]
\end{equation}
where $B_1 = [-3 \sqrt{2} \pi ^2+8 \left(-1+2 \sqrt{2}\right) \sqrt{\pi } \zeta(\frac{3}{2})]/64 \approx 0.4$ and $\zeta(x)$ is the zeta function.
Then the  inverse  compressibility is given by
\begin{equation}
\frac{K_0}{K(T)} = 1 -
\frac{r_s}{\pi}\left(\frac{B_1}{\sqrt{t}}+ \frac{p(y)-q(y)}{3}-w(y) \right).
\end{equation}
In the  high temperature limit ($t \gg 1$)
the inverse compressibility of the ordinary 2D systems increases linearly with
temperature and approaches the non-interacting value. However,
since the non-interacting compressibility of BLG is temperature independent
and the most dominant exchange term decreases as $t^{-1/2}$, the high temperature
inverse compressibility of BLG approaches the following high-temperature limit:
\begin{equation}
\frac{K_0}{K(T)}\Big|_{T\rightarrow \infty} = 1 - \frac{1}{8}\frac{a}{a_B},
\end{equation}
where $a_B= \hbar^2\kappa/(me^2)$ is the BLG effective Bohr
radius. Since $a/a_B <1$ the negative inverse compressibility reverses
its sign to the positive values at high temperatures and asymptotically
approaches a smaller value than the zero temperature inverse compressibility. This intriguing result is purely a  high-temperature manifestation of exchange effect within the 2-band BLG approximation.

In Fig.~\ref{fig:blg}, we present our numerical results of temperature
dependent inverse compressibility in BLG systems. In the calculation we use
$k_c=1/a$ ($a=2.46${\AA}) and $\kappa=2.5$.
As shown in Fig.~\ref{fig:blg} the calculated BLG inverse compressibility shows
very weak temperature dependence. The weak non-monotonic behavior of
temperature-dependent inverse compressibility is shown in the inset of
Fig.~\ref{fig:blg}(a). In
Fig.~\ref{fig:blg}(b), we present the inverse compressibility versus
$r_s$. $K^{-1}$ decreases monotonically with $r_s$ and becomes negative at
$r_s \sim 27$.
In Fig.~\ref{fig:blg}(c)) we show $\partial \mu /\partial n$ as a
function of density and the inset shows the same figure in the
semi-logarithm scale to show clearly the behavior at low densities. At
densities $n \alt 10^{10} cm^{-2}$ the zero temperature $\partial \mu
/\partial n$ is negative, but as temperature increases it reverses its
sign at $t \approx [0.4/(\pi/r_s-a/8a_B)]^2$. As temperature
increases further it approaches $(\hbar^2 \pi/2m)(1-a/8a_B)$ regardless of density.

\begin{figure}
\begin{center}
\includegraphics[width=0.7\linewidth]{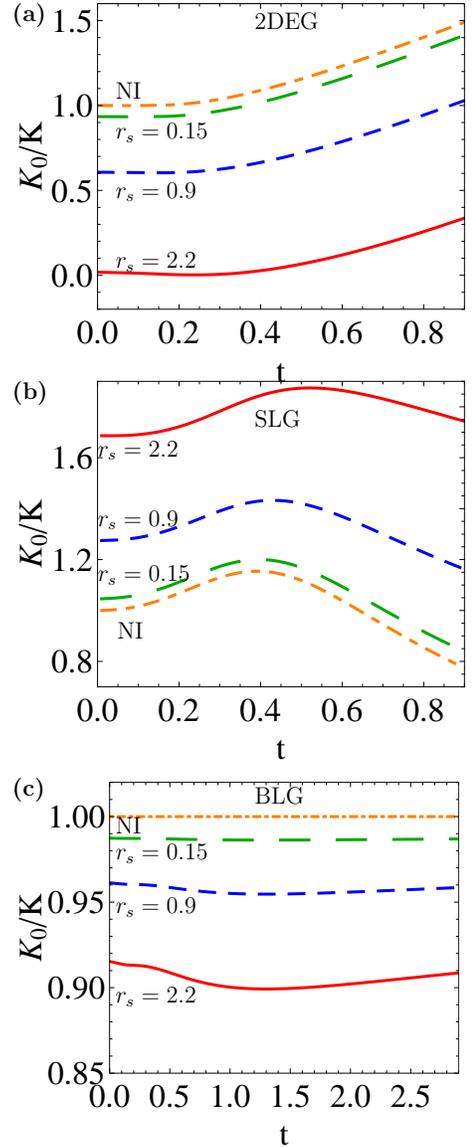}
\caption{(color online) Calculated $K_0/K$ for the 2D, MLG, and BLG systems as a function of $t$ for a few values of $r_s$.}
  \label{fig:blgF}
\end{center}
\end{figure}

In Fig.~\ref{fig:blgF}, we show, for the purpose of comparison, the calculated $K_0/K$ for the 2D, MLG, and BLG systems as functions of $r_s$ and $t$ in order to demonstrate the qualitative difference between the results.

\section{Discussion and Conclusion}  \label{sec:discussion}

In this work, we have calculated the finite temperature compressibility of
graphene and ordinary 2D semiconductor systems within the Hartree-Fock
approximation. We present both analytical and numerical results of
$K_0/K$ as a function of temperature, the dimensionless interaction parameter
$r_s$, and density. We find that the calculated temperature dependent
compressibility is non-monotonic in both graphene and  2D
semiconductor systems. In monolayer graphene, the inverse compressibility
$K^{-1}$ increases with temperature at low temperatures and decreases
at high temperatures, reaching a maximum value at an intermediate
temperature. This non-monotonicity arises entirely from the behavior of the non-interacting graphene compressibility. The temperature dependent inverse
compressibility in 2D semiconductor systems decreases with
temperature at low temperatures and increases at high temperatures. In
BLG the inverse compressibility decreases as temperature increases at
low temperatures, but it approaches a value which is less than the
non-interacting value at high temperatures. The
leading order temperature correction to the  inverse compressibility $K^{-1}$ in
HFA for both graphene and 2D semiconductor systems is $t^2 \ln
t$. Our analytic results are in agreement with our numerical results
both at low and at high temperatures.

Our use of the Hartree-Fock approximation is not a particularly
restrictive approximation for the theory of the electronic
compressibility since it is well-known that Hartree-Fock theory works
well for the calculation of the compressibility and its density
dependence, at least at $T=0$.  The Hartree-Fock approximation gives
results numerically very close to the full dynamical RPA
for the 2D compressibility at $T=0$
\cite{DasSarmaHwang_PRB09}, and the Hartree-Fock compressibility
results are in good agreement with the compressibility measurements
for 2D semiconductors\cite{eisenstein:1994} and
graphene\cite{Euyheon_PRL07}.  In fact, the Hartree-Fock approximation
for 2D semiconductors is in better quantitative agreement with the
density dependence of the measured
compressibility\cite{eisenstein:1994}  than the corresponding RPA
theory as long as the strict 2D approximation is used since the finite
width corrections for the realistic 2D semiconductors tend to cancel
out the correlation corrections neglected in the HFA.  In any case, we
consider the HFA as the first step necessary for understanding the
temperature dependent 2D compressibility in interacting electron
systems, and only future work, particularly experiments comparing with
our predictions, can establish the necessity of improved theoretical
treatments involving dynamical correlations neglected in the HFA.

Our main finding that the temperature dependence of compressibility is
much stronger in 2D systems than in graphene is consistent with
existing experimental results. In particular, our results of
Sec.~\ref{sec:semi} agree well with the experimental study of Shapira
{\it et al.},\cite{shapira:1996} on 2D GaAs holes who
discovered substantial temperature dependence in the 2D hole
compressibility in the $1-100$ K temperature regime with the
experimental data showing reasonable qualitative agreement with our
theoretical HFA results presented in Sec.~\ref{sec:semi}D. Detailed quantitative comparison with the
experimental data of Shapira {\it et al.,}\cite{shapira:1996} is not
particularly meaningful (and
is not attempted) because of a number of reasons including our use of
the exchange-only Hartree-Fock approximation, our neglect of the
finite thickness of the 2D semiconductor structures which is often
important for quantitative considerations, our neglect of disorder
effects which are certainly important at lower carrier densities, and
the lack of information about some essential experimental parameters
(e.g., depletion charge density).

In addition to the experimental study of the temperature dependent 2D
hole compressibility by Shapira {\it et al.,} \cite{shapira:1996}
discussed above, there have been
several low-density studies of the temperature dependence of the
compressibility in both electron and hole 2D
systems\cite{eisenstein:1992,eisenstein:1994,shapira:1996}
carried out in the context of 2D MIT (i.e., disorder-induced
density-tuned localization of 2D semiconductor systems). As we
discussed in Sec.~\ref{sec:semi}, these low-density experimental
investigations should be revisited in light of our current work
demonstrating the importance of the temperature dependence of $K_0/K$
in the low carrier density regime by virtue of the dimensionless
temperature $T/T_F$ being large precisely in the low-density regime
(where disorder effects are also strong) since $T_F \propto n$ in 2D
systems. Unfortunately, temperature, disorder, and interaction effects
are all strong in the low density regime, making any quantitatively
reliable theoretical work essentially impossible in the low carrier
density regime, and our work establishes the importance of finite
temperature effects on the compressibility in the low carrier density
regime. In particular, the density inhomogeneity and puddle formation
that happens at low carrier density
\cite{yamaguchi_2008_new,Allison_PRL06} due to the failure of
screening would very much complicate the observation of the
low-density (and low-temperature) features in the compressibility
predicted in our theory unless one uses extremely high-quality samples
with very low disorder.
Recently, there has been a theoretical investigation
\cite{DavidHwang_PRB11} of the effect of disorder on low-density
graphene compressibility clearly establishing the importance of
disorder in low-density compressibility.

Since the Fermi temperature $T_F$ goes as $T_F$(in K)$=4.2 \tilde{n}$ (2D GaAs electrons); $0.74 \tilde{n}$ (2D GaAs holes); $136 \sqrt{\tilde{n}}$ (monolayer graphene); $4.2 \tilde{n}$ (bilayer graphene), where $\tilde{n} \equiv n/10^{10}$ is the carrier density measured in the units of $10^{10}$ cm$^{-2}$, it is obvious that the quantitative effect of finite temperature, even at low carrier densities, is by far the strongest (weakest) in 2D GaAs holes (monolayer graphene), which is consistent with experimental observations. Since disorder effects become important for $n \lesssim n_i$, where $n_i$ is the background random charged impurity density in the environment, the low density regime associated with large $T/T_F$ values becomes even more challenging to achieve in the laboratory experiments on graphene. We expect our predicted temperature dependence to manifest in graphene compressibility in room temperature experiments in very high-mobility samples where $T/T_F \sim 1$ values may be achievable.

An important relevant question for our theory is: what is the most
suitable system for the experimental observation of our theoretical
predictions? Obviously, our predicted high-temperature behavior, where
the theory is in very firm ground since the exchange energy is likely
to be the exact leading-order many-body correction to the
non-interacting compressibility in the high-temperature limit, should
be valid for all systems and should be observable in 2D semiconductors
(both electrons and holes) rather easily by measuring the
compressibility for $T=10-200$ K for electron (hole) densities$\sim
10^{10}-10^{11}$ cm$^{-2}$ so that $T>r_s^2 T_F$ condition is
satisfied. Given that $T_F^{h}<T_F^{e}$ at the same density,
the high-temperature behavior is much more easily observable in 2D
GaAs holes, and has, in fact, already been observed at least in one
experimental study in 2D holes\cite{shapira:1996}. While
2D GaAs holes are the obvious ideal candidates for observing our
predicted high-temperature compressibility behavior, monolayer
graphene may turn out to be not particularly well-suited for the
temperature-dependent compressibility studies because of its very high
relative Fermi-temperature ($T_F^{MLG} \sim 1350$ K for a doping
density of $10^{12}$ cm$^{-2}$) and its relatively weak temperature
dependence. Obviously, the condition for observing the
high-temperature behavior of compressibility is the ability to reach
$T \gtrsim T_F$ which necessitates lower Fermi temperatures. By
contrast, the low-$T$ behavior associated with the $T^2 \ln T$ term in
the compressibility may be better observed in a system with a
relatively high value of $T_F$ so that the regime $T/T_F <1 $ can be
explored over a fairly broad range of temperature with the temperature
$t_m$ not being too low. Thus, very high mobility n-GaAs
heterostructures or quantum wells with $n \sim 2 \times 10^{10}$ cm$^{-2}$ so
that $T_F \sim 8.3$ K (and $T_m \sim 3$ K) may be the ideal system to
search for our predicted shallow minimum in $K_0/K$. One serious
problem is that the minimum may be too shallow to be uniquely
determined experimentally.

Finally, we discuss our very interesting analytical finding of the
non-monotonicity in graphene inverse compressibility associated with the $T^2
\ln T$ correction we obtain analytically (and verify
numerically). Although this would not be an easy effect to detect
experimentally because of its quantitative weakness (and became the
subleading correction goes as $T^2$ and thus the ratio of the two,
$\ln (T/T_F)$, is always challenging to observe in experiments even
under the best of circumstances), it is nevertheless interesting to
discuss its origin and its robustness beyond our approximation
scheme. Our results are exact within the Hartree-Fock approximation
scheme, and there is no doubt that the exchange self-energy
contributes a $T^2 \ln T$ leading-order contribution to the inverse
compressibility. The first question is whether this is intrinsically a
dimensionality effect occurring only in  two dimensions. We have
therefore carried out the corresponding analytical calculations for
the 3D temperature-dependent compressibility (presented in the
Appendix \ref{sec:appen} of this paper), finding that $K_0/K$ has a $T^2 \ln T$
contribution in three dimensions also. Thus, it appears that the
interesting non-monotonicity associated with the $T^2 \ln T$
correction is a results of the Hartree-Fock approximation arising from
the exchange self-energy diagram, and is not intrinsically a
two-dimensional effect.

We have investigated this question by
calculating the interacting compressibility in the screened
Hartree-Fock approximation (sometimes also called static RPA) for the
2D system, where the bare Coulomb interaction `$V$' appearing in the
exchange self-energy is screened by the static RPA dielectric
function. (These results are shown in Appendix \ref{sec:appenB}.) The
screened HFA (Appendix \ref{sec:appenB}) or static RPA does not
contain the $T^2 \ln T$ term in the finite temperature inverse
compressibility, but is not a reliable approximation at all since at
$T=0$ it predicts that $K_0/K$ should be always positive for all
values of $r_s$ in clear disagreement with experimental
finding\cite{eisenstein:1992,eisenstein:1994}. Thus the absence of the
$T^2 \ln T$ term in the screened HF approximation cannot be taken
seriously since the corresponding $T=0$ result is in qualitative
disagreement with experimental results.  It is interesting to
speculate whether the
$T^2 \ln T$ term survives higher-order diagrams associated with
dynamical correlation effects, and this remains a challenge for the
future. Although it is fairly straightforward to calculate the
compressibility including correlation effects at $T=0$ --- for
example, the ring-diagram contributions to the compressibility can be
exactly calculated at $T=0$\; \cite{DasSarmaHwang_PRB09} --- it is a
formidable challenge, both numerically and analytically, to extend to
the corresponding ring diagram calculations of interacting
compressibility to finite temperatures. We have studied this question
carefully and have been able to show that there is definitely a $T^2$
term in $K_0/K$ arising from the infinite series of the ring diagrams,
but we still do not know whether the $T^2 \ln T$ term of the exchange
self-energy is exactly canceled by the dynamical effects arising from
the infinite ring diagram series. Although the effect of the
$(T/T_F)^2 \ln (T/T_F)$ term matters only at low temperatures and low
densities, where disorder effects dominate making it difficult to
observe the log term experimentally, it is nevertheless important to
establish whether this non-monotonicity associated with $T^2 \ln T$
survives higher-order correlation terms in the theory. We leave this
as an unanswered theoretical question for the future. A direct
experimental observation of the $T^2 \ln T$ term in the
low-temperature inverse compressibility of 2D electrons or holes will go a
long way in settling this important question.

To summarize, our goal of this paper
is to understand the temperature dependence of compressibility in
a high density homogeneous system where the interaction effect is not
too strong and HFA is valid. At very low density, disorder effects are of particular importance
 and the system may be highly
inhomogeneous due to the formation of electron-hole puddles\cite{ilani:2000,DultzJiang_arXiv,shi:2002,DasSarma_RMP11,gao:2011,DavidHwang_PRB11,DMHD_arXiv11,yamaguchi_2008_new,Allison_PRL06,hwang:2010,li:2011,RossiDas_PRL08}, which are not included in our theory. Our main result is that temperature effects in the compressibility could be quite important at higher temperatures.

\begin{acknowledgments}
The work is supported by US-ONR-MURI and NRI- NSF-SWAN.

\end{acknowledgments}

\appendix
\section{Hartree-Fock compressibility in 3DEG }
\label{sec:appen}

In this Appendix, we provide the corresponding finite
temperature Hartree-Fock compressibility for a standard three
dimensional electron gas (3DEG). For 3DEG, the bare Coulomb
interaction $V(q)=\dfrac{4\pi e^2}{\kappa q^2}$, where $\kappa$ is the
background dielectric constant. We assume the total degeneracy $g=2$
in the calculation.

The finite temperature exchange energy of 3DEG is given by
\begin{eqnarray}
\begin{array}{l l l l l }
\Sigma _{\text{ex}}(k)&=&-\int \frac{d^3k'}{(2\pi )^3}V(|k-k'|)n_F(\xi _{k'})
\\
\\
&=&- \dfrac{e^2 k}{\pi  \kappa }\int _0^1 \text{d}x \ln\left(\dfrac{1+x}{1-x}\right)\Big[\dfrac{x}{e^{\beta (\epsilon _k x^2-\mu _0)}+1}
\\
\\
&+&\dfrac{1/x^3}{e^{\beta (\epsilon _k/x^2-\mu _0)}+1}\Big]
\end{array}
\end{eqnarray}
where $\beta=1/k_BT$, $\epsilon _k=k^2/2m$, $m$ is the effective mass,
and $\mu _0$ is the finite temperature chemical potential without
exchange energy. The non-interacting chemical potential, $\mu_0$, is obtained by
solving the following self-consistent equation,
\begin{equation}
n=\dfrac{k_F^3}{ \pi ^2}\int _0^{\infty }\text{d} x\frac{x^2}{\exp(\frac{x^2-\tilde{\mu_0}}{t})+1}
\end{equation}
where $\tilde{\mu_0}=\mu_0/E_F$, $t=T/T_F$ and $n=\dfrac{k_F^3}{3
  \pi^2}$ for 3DEG. There is no explicit analytical formula for the
finite temperature chemical potential in 3DEG. But at very low
temperatures ($t \ll 1$), the asymptotic form of the non-interacting
chemical potential is given by
\begin{eqnarray}
\begin{array}{l l}
\mu _0 \approx E_F \left(1 - \dfrac{\pi^2 t^2}{12}\right)
\end{array}
\end{eqnarray}

The HFA chemical potential, $\mu$, can be calculated by including the
exchange energy for $k=k_F$, i.e.,
\begin{equation}
\mu = \mu_0(T) +\Sigma_{ex}(k_F,T).
\label{eq:chemi_a}
\end{equation}
Then, the normalized chemical potential $\tilde{\mu}=\mu/E_F$ can be expressed
\begin{eqnarray}
\tilde{\mu}=\tilde{\mu_0}-\frac{2 \alpha}{\pi} \int _0^1\text{d}x
\ln \left ( \dfrac{1+x}{1-x} \right ) \left [ \frac{x}{e^{(x^2-\tilde{\mu
        _0})/t}+1} \right . \nonumber \\
 \left . +\frac{1/x^3}{e^{(1/x^2-\tilde{\mu _0})/t}+1} \right ],
\label{eq:chemire_a}
\end{eqnarray}
where $\alpha=({4}/{9\pi})^{1/3}r_s$ with the dimensionless parameter
$r_s = (3/4\pi n)^{1/3}(me^2/\kappa)$, and
\begin{equation}
\tilde{\mu_0}=1 - \frac{\pi^2 t^2}{12}.
\end{equation}
By differentiating Eq.~(\ref{eq:chemi_a}) with respect to $n$ and
using $k_F^3=3 \pi^2 n$, the corresponding non-interacting
compressibility of 3DEG is given by
\begin{eqnarray}
\frac{1}{n^2 K_0}=\frac{d E_F}{d n}=\frac{\pi^2}{ m k_F}
\\
\frac{1}{n^2 K_{ni}}=\frac{d \mu_0}{d n}=\frac{h_3(t)}{K_0},
\end{eqnarray}
where $K_0$ and $K_{ni}$ are the zero temperature and the finite
temperature non-interacting compressibility, respectively, and
$h_3(t)=1+ \dfrac{\pi^2 t^2}{12}$.

The finite temperature inverse compressibility $K_0/K$ within HFA is given by
\begin{widetext}
\begin{eqnarray}
\frac{K_0}{K}=h_3(t) & - &\dfrac{\alpha}{\pi} \int _0^1\text{d}x
\ln \left ( \dfrac{1+x}{1-x} \right ) \left [
  \frac{x}{e^{(x^2-\tilde{\mu _0})/t}+1} +
  \frac{1/x^3}{e^{(1/x^2-\tilde{\mu _0})/t}+1} \right ] \nonumber
\\
& + & \dfrac{\alpha}{2\pi t}  \int _0^1\text{d}x \ln \left (
\dfrac{1+x}{1-x} \right ) \left [ \frac{x (x^2-h(t))}{\cosh
    ^2\frac{(x^2-\tilde{\mu _0})}{2t}}
  +\frac{\frac{1}{x^3}(\frac{1}{x^2}-h(t))}{\cosh
    ^2\frac{(1/x^2-\tilde{\mu _0})}{2t}} \right ].
\label{eq:comre_a}
\end{eqnarray}
\end{widetext}
At zero temperature $T=0$, we have
\begin{equation}
\frac{K_0}{K} = 1-\frac{\alpha}{\pi}=
1-\frac{1}{\pi}\sqrt[3]{\dfrac{4}{9\pi}}r_s.
\end{equation}
The HF chemical potential and the compressibility in the low
temperature limit $t \ll 1$ can be calculated by expanding the
logarithmic function near $x=1$.  Then the chemical potential in
Eq.~(\ref{eq:chemire_a}) becomes
\begin{equation}
\tilde{\mu}=\tilde{\mu_0}-\frac{2}{\pi}\sqrt[3]{\dfrac{4}{9\pi}}r_s
\left [1+\frac{\pi^2 t^2}{24}( A_3 +\text{ln}t) \right ],
\end{equation}
where $A_3=1-\gamma+\text{ln}\Big(\dfrac{\pi}{8}\Big)$ (where $\gamma$
is Euler's constant, with numerical value $\simeq 0.577216$.). Finally
we find that the low temperature 3D inverse
compressibility within HFA as
\begin{eqnarray}
\dfrac{K_0}{K}\Big|_{t\rightarrow 0} & = & 1+\dfrac{\pi ^2 t^2}{12} -
\sqrt[3]{\dfrac{4}{9\pi^4}}r_s \left [ 1 \right . \nonumber \\
& + &  \left . 0.136\pi ^2 t^2
  +\dfrac{1}{8} \pi ^2 t^2\text{  }\ln (\dfrac{\pi  t}{8}) \right ].
\end{eqnarray}
The asymptotic behavior of high temperature compressibility is
obtained by approximating the Fermi-Dirac distribution function by the
classical Boltzmann function. In the high temperature regime $t
\rightarrow \infty$, the normalized chemical potential within
HFA becomes
\begin{equation}
\tilde{\mu}\Big|_{t \rightarrow \infty} = t \ln \left ( \frac{4}{3
  \sqrt{\pi} t^{3/2}} \right ) - \frac{8}{3 t}\sqrt[3]{\frac{4}{9\pi^4}}r_s,
\end{equation}
and the corresponding high temperature inverse compressibility is calculated as
\begin{equation}
\dfrac{K_0}{K}\Big|_{t\rightarrow \infty}=\frac{3 t}{2}-\frac{4}{t}\sqrt[3]{\frac{4}{9\pi^4}}r_s,
\end{equation}
which agrees with the non-interacting high-temperature classical
result.

\section{Screened 2D Hartree-Fock compressibility}
\label{sec:appenB}

We provide the screened Hartree-Fock compressibility (sometimes also called static RPA) for the 2D system, where the bare Coulomb interaction `$V$' appearing in the exchange self-energy is screened by the static RPA dielectric function.

The self-energy  with the finite temperature static RPA dielectric
function is given by
\begin{equation}
\Sigma_{\text{self}}(k_F)=-\int \frac{d^2q}{(2\pi )^2}\frac{V(q)}{\epsilon (q,T)} n_F\left(\xi _{q-k_F}\right)
\label{eq:selfB}
\end{equation}
where $V(q)=\dfrac{2\pi  e^2}{\kappa q}$, $n_F\left(\xi
_{q-k_F}\right)$ is the Fermi distribution function. The static RPA
dielectric function $\epsilon\left(q,T\right)$ is given by \cite{DasSarma_RMP11}
\begin{equation}
\epsilon\left(q,T\right) = 1 + V(q)\Pi(q,T)= 1+  \frac{q_s(q,T)}{q}
\label{eq:apBepsi} %
 \end{equation}
 where $\Pi(q,T)$ is the finite temperature static polarization
 function and $q_s=2\pi e^2 \Pi(q,T)/\kappa$ is the momentum dependent
 screening wave vector \cite{DasSarma_RMP11}.
At low temperatures ($T\ll T_F$) the 2D
polarizability becomes \cite{DasSarma_RMP11},
\begin{equation}
\Pi(q,T)\approx \frac{m}{\pi} \left(1-e^{-T_F/T}\right),
\label{eq:apBlowT}
\end{equation}
and its asymptotic form at high temperatures ($T \gg T_F$)
becomes\cite{DasSarma_RMP11}
\begin{equation}
\Pi(q,T) \approx \frac{m }{\pi }\frac{T_F}{T}.
\label{eq:apBhighT}
\end{equation}

The finite temperature chemical potential within static RPA, $\mu(T)$,
can be calculated by including the self-energy of
Eq.~(\ref{eq:selfB}), i.e.,
\begin{equation}
\mu(T) = \mu_0(T) +\Sigma_{\text{self}}(k_F,T).
\label{eq:chemi_B}
\end{equation}
At zero temperature ($T=0$), the non-interacting part of chemical
potential is the Fermi energy $\mu = E_F$. For $r_s
\ll 1$, the chemical potential approaches the non-interacting
value, which has been discussed in Sec.~\ref{sec:semi}. For
interacting 2D systems the self-energy within screened HFA has the
asymptotic form (with $g_s=2$ and $g_v=1$) for $r_s \gg 1$,
\begin{equation}
\Sigma_{\text{self}}(k_F,T=0)\Big|_{r_s \gg 1}= - \frac{\pi n}{2 m }.
\label{eq:self0}
\end{equation}
Differentiating Eq. \ref{eq:self0} with respect to $n$ and using
the relation $k_F^2 = 2 \pi n$, we get the zero temperature  inverse compressibility within the screened HFA
\begin{equation}
\frac{K_0}{K(T=0)}\Big|_{r_s \gg 1} = \frac{1}{2}.
\label{eq:apcom0}
\end{equation}
Since $K_0$ is a positive value we have the positive
compressibility even for $r_s \gg1$, which
clearly disagrees with both the HF results and
experiments\cite{eisenstein:1992,eisenstein:1994}.

With Eqs.~(\ref{eq:selfB})--(\ref{eq:apBlowT}), we find  the
asymptotic form of the self-energy at low temperatures ($T \ll T_F$) and
for $r_s \gg 1$
\begin{equation}
\Sigma_{\text{self}}(k_F,T)\Big|_{r_s \gg 1}=  - \frac{\pi n}{2 m } + O(e^{-T_F/T}).
\end{equation}
We find that the leading order temperature dependent term in the self
energy is exponentially suppressed. Consequently,
the temperature dependent inverse compressibility at low temperatures becomes
\begin{equation}
\frac{K_0}{K(T \ll T_F)}\Big|_{r_s \gg 1} \approx \frac{1}{2} + O(e^{-T_F/T}).
\label{eq:apcomT}
\end{equation}
At high temperatures ($T \gg T_F$), the inverse
compressibility calculated within the static RPA approaches the HFA
results given in Eq.~(\ref{eq:semiht}) because the non-interacting
kinetic energy dominates.

\end{document}